\documentclass[12pt, draftclsnofoot, onecolumn]{IEEEtran}
\usepackage{amsthm}
\usepackage{multirow}
\usepackage{color}
\newtheorem*{lemma1}{Lemma 1}

\usepackage[T1]{fontenc}% optional T1 font encoding
\usepackage{algorithm}
\usepackage{algorithmic}
\usepackage{multirow}
\usepackage{longtable}

\ifCLASSINFOpdf
\else
\fi
\usepackage{amsmath}
\usepackage{amsmath,bm}
\usepackage{graphicx}
\usepackage{amsmath,amssymb,amsthm,mathrsfs,amsfonts,dsfont}
\usepackage{subfigure}
\usepackage{array}
\begin{document}
% Titles are generally capitalized except for words such as a, an, and, as,(38)
\title{Directional Frame Timing Synchronization in Wideband Millimeter-Wave Systems with Low-Resolution ADCs}
%ft

\author{Dalin Zhu,
        Ralf Bendlin,
        Salam Akoum,
        Arunabha Ghosh,
        and~Robert~W.~Heath,~Jr.% <-this % stops a space
\thanks{Dalin Zhu and Robert W. Heath, Jr. are with the Department
of Electrical and Computer Engineering, The University of Texas at Austin, Austin,
TX, 78712 USA, e-mail: \{dalin.zhu, rheath\}@utexas.edu.

Ralf Bendlin, Salam Akoum and Arunabha Ghosh are with AT\&T Labs, Austin, TX, 78759 USA, e-mail: \{ralf\_bendlin, salam\_akoum, ghosh\}@labs.att.com.

Parts of this work have been presented at the IEEE Fifty-First Asilomar Conference on Signals, Systems and Computers \cite{dzasilomar}. This work was supported in part by the National Science Foundation under Grant No. ECCS-1711702, CNS-1702800 and CNS-1731658 and a gift from AT\&T Labs.}}

\maketitle


\begin{abstract}
In this paper, we propose and evaluate a novel beamforming strategy for directional frame timing synchronization in wideband millimeter-wave (mmWave) systems operating with low-resolution analog-to-digital converters (ADCs). In the employed system model, we assume multiple radio frequency chains equipped at the base station to simultaneously form multiple synchronization beams in the analog domain. We formulate the corresponding directional frame timing synchronization problem as a max-min multicast beamforming problem under low-resolution quantization. We first show that the formulated problem cannot be effectively solved by conventional single-stream beamforming based approaches due to large quantization loss and limited beam codebook resolution. We then develop a new multi-beam probing based directional synchronization strategy, targeting at maximizing the minimum received synchronization signal-to-quantization-plus-noise ratio (SQNR) among all users. Leveraging a common synchronization signal structure design, the proposed approach synthesizes an effective composite beam from the simultaneously probed beams to better trade off the beamforming gain and the quantization distortion. Numerical results reveal that for wideband mmWave systems with low-resolution ADCs, the timing synchronization performance of our proposed method outperforms the existing approaches due to the improvement in the received synchronization SQNR.
\end{abstract}

%\begin{IEEEkeywords}
%Auxiliary beam pair, millimeter-wave (mmWave), channel estimation, analog-only beamforming, hybrid analog and digital precoding, angle-of-departure (AoD), angle-of-arrival (AoA).
%\end{IEEEkeywords}
% \fi
\IEEEpeerreviewmaketitle

\allowdisplaybreaks

\section{Introduction}
The millimeter-wave (mmWave) band offers high data rates in both wireless local area networks \cite{jwzl,ieeewlan} and fifth-generation (5G) mobile cellular systems \cite{rhsp,5gnr}. As mmWave systems make use of large available bandwidths, the corresponding sampling rate of the analog-to-digital converters (ADCs) scales up, which results in high power consumption and hardware implementation complexity. It is desirable to reduce the ADCs' resolution in mmWave systems to reduce implementation costs \cite{hpadc}. The use of low-resolution ADCs in wireless communications systems has been investigated in various aspects, e.g, input signal optimization in \cite{amjn}-\nocite{jsodum}\cite{amjan} and mutual information analysis in \cite{jmrwj}-\nocite{mijn}\cite{bmic}. In practice, low-resolution ADCs will also impair the frame timing synchronization performance of mmWave systems \cite{awumw}. That issue, however, was not addressed in \cite{amjn}-\nocite{jsodum}\nocite{amjan}\nocite{jmrwj}\nocite{mijn}\cite{bmic}. In summary, most of the prior work on low-resolution ADCs focused on analytical performance assessment rather than practical implementation issues such as synchronization.
%, channel estimation techniques in \cite{tlvw}-\nocite{gzgk}\cite{jmps}, and uplink multiuser detection algorithms in \cite{swyl}-\nocite{jcjmrj}\cite{cmjcel}; \nocite{tlvw}\nocite{gzgk}\nocite{jmps}\nocite{swyl}\nocite{jcjmrj}\cite{cmjcel}

Current lower-frequency cellular networks such as long-term evolution (LTE) systems \cite{lte} conduct frame timing synchronization using omni-directional transmission and reception. Directional transmission and detection of synchronization signals is interesting in mmWave systems due to the low signal-to-noise ratio (SNR) prior to beamforming \cite{wonil}. For directional synchronization, the network sends synchronization signals towards predefined angular directions via beamforming \cite{5gnr,5gbmgnt}. Most of the prior work on beamforming based directional synchronization design, however, did not incorporate the effect of low-resolution quantization.

In this paper, we propose and evaluate a new beamforming strategy to improve the frame timing synchronization performance for mmWave cellular systems under low-resolution ADCs. Different from our prior work in \cite{dzasilomar}, in which the developed synchronization method mainly focused on a single user setup, the proposed approach in this paper incorporates multiple users and optimizes the overall synchronization performance. In our system model, the base station (BS) deploys multiple radio frequency (RF) chains and simultaneously forms multiple synchronization beams in the analog domain. Upon receiving the synchronization signals, the user equipment (UE) conducts cross-correlation based frame timing synchronization with fully digital front ends and low-resolution ADCs. We summarize the main contributions of the paper as follows:
\begin{itemize}
  \item \emph{Optimization problem formulation for directional frame timing synchronization under low-resolution ADCs}: For a single UE with low-resolution ADCs, we leverage Bussgang's decomposition theorem \cite{buss1,buss2} to formulate the corresponding received synchronization signal-to-quantization-plus-noise ratio (SQNR) at zero-lag correlation. This formulation accounts for both the spatial correlation brought by the directional beamforming and the inherent correlation of the employed synchronization signals. Building on the derivation of a single UE's synchronization SQNR, we extend the problem formulation of low-resolution synchronization to a multi-user scenario. In this case, we focus on maximizing the minimum received synchronization SQNR at zero-lag correlation among all UEs. We show that this type of max-min multicast problem cannot be effectively solved by existing single-stream beamforming based approaches due to large quantization distortion and limited beam codebook resolution.
  \item \emph{New directional frame timing synchronization design under low-resolution ADCs}: Without channel knowledge (a common assumption for synchronization), we first discretize the given angular range with a set of potential channel directions and transform the complex max-min multicast problem into a maximization problem. We then develop a new multi-beam probing based directional synchronization strategy to tackle this problem. Leveraging a common synchronization signal structure design, the simultaneously probed synchronization beams form an effective composite beam. We show that by optimizing the effective composite beam pattern, a good tradeoff between the beamforming gain and the resulted quantization distortion can be achieved, resulting in improved frame timing synchronization performance under low-resolution quantization.
  %By discretizing the given angular range with a set of potential channel directions, we transform the complex max-min multicast problem to a simple maximization problem, which depends on the selection of the beamforming vectors. By exploiting the spatial design degrees of freedom brought by the multi-beam probing, we optimize the combination of the simultaneously probed synchronization beams to tackle this problem. Leveraging a common synchronization signal structure, the simultaneously probed beams result in an effective composite beam that not only increases the received synchronization SQNR at zero-lag correlation, but also characterizes the worst-case scenario of the network.
\end{itemize}
In essence, optimizing the received synchronization SQNR at zero-lag correlation is a viable solution to improve the overall frame timing synchronization performance under low-resolution ADCs. This is because for well-structured synchronization signals, the non-zero-lag correlation values are small and barely affected by the quantization, while the zero-lag peak correlation value is significantly distorted by the quantization. In Sections III and V, we use several analytical and numerical examples to reveal these observations. We also conduct simulations of wideband mmWave cellular systems, showing that the proposed design approach can achieve promising received synchronization SQNR, frame timing position estimation, and access delay performances assuming low-resolution quantization.

We organize the rest of the paper as follows. In Section II, we specify the system and channel models for the directional frame timing synchronization design in mmWave systems. In Section III, we formulate the directional frame timing synchronization problem under low-resolution ADCs. In Section IV, we explicitly illustrate the design principle and procedure of the proposed multi-beam probing strategy. We evaluate the proposed synchronization method in Section V assuming both narrowband and wideband channels. We draw our conclusions in Section VI.

\textbf{Notations}: $\bm{A}$ ($\textbf{\textsf{A}}$) is a matrix; $\bm{a}$ ($\textbf{\textsf{a}}$) is a vector; $a$ ($\textsf{a}$) is a scalar; $|a|$ is the magnitude of the complex number
$a$; $(\cdot)^{\mathrm{T}}$ and $(\cdot)^{*}$ denote transpose and conjugate transpose; $((\cdot))_{N}$ represents the modulo-$N$ operation; $\left[\bm{A}\right]_{:,j}$ is the $j$-th column of $\bm{A}$; $\left[\bm{A}\right]_{i,j}$ is the $(i,j)$-th entry of $\bm{A}$; $\left[\bm{a}\right]_{j}$ represents the $j$-th element of $\bm{a}$; $\left[\bm{a}\right]_{j_1:j_2}$ contains elements $j_1,j_1+1,\cdots,j_2$ of $\bm{a}$; $\left[\bm{A}\right]_{:,j_1:j_2}$ contains columns $j_1,j_1+1,\cdots,j_2$ of $\bm{A}$; $\mathrm{tr}(\bm{A})$ is the trace of $\bm{A}$; $\bm{I}_{N}$ is the $N\times N$ identity matrix; $\bm{1}_{M\times N}$ represents the $M\times N$ matrix whose entries are all ones; $\bm{0}_{N}$ denotes the $N\times 1$ vector whose entries are all zeros; $\mathcal{N}_{c}(\bm{a},\bm{A})$ is a complex Gaussian vector with mean $\bm{a}$ and covariance $\bm{A}$; $\mathbb{E}[\cdot]$ is used to denote expectation; the diagonal matrix $\mathrm{diag}(\bm{A})$ has $\left\{\left[\bm{A}\right]_{i,i}\right\}$ as its diagonal entries, and $\mathrm{diag}(\bm{a}^{\mathrm{T}})$ has $\left\{\left[\bm{a}\right]_{j}\right\}$ as its diagonal entries; and $\mathrm{diag}\left\{\cdot\right\}$ is the diagonalization operation.

\section{System and Channel Models}
In this section, we introduce the system model for the directional frame timing synchronization design in mmWave systems. We also summarize the common wideband channel model.
\subsection{System model for directional frame timing synchronization in mmWave systems}
In this paper, we assume that the BS employs directional beams to transmit the downlink synchronization signals, providing sufficient link margin at mmWave frequencies. Note that the directional transmission of downlink synchronization signals will also be supported in the 3GPP 5G New Radio (NR) systems \cite{5gnr,5gbmgnt}. In the following, we first present the assumed antenna array configurations and transceiver architecture along with the synchronization signal structure. We then develop the received synchronization signal model for our system. Finally, we explain the cross-correlation based frame timing synchronization design using the derived received signal model.

\begin{figure}
\centering
\subfigure[]{%
\includegraphics[width=4.05in]{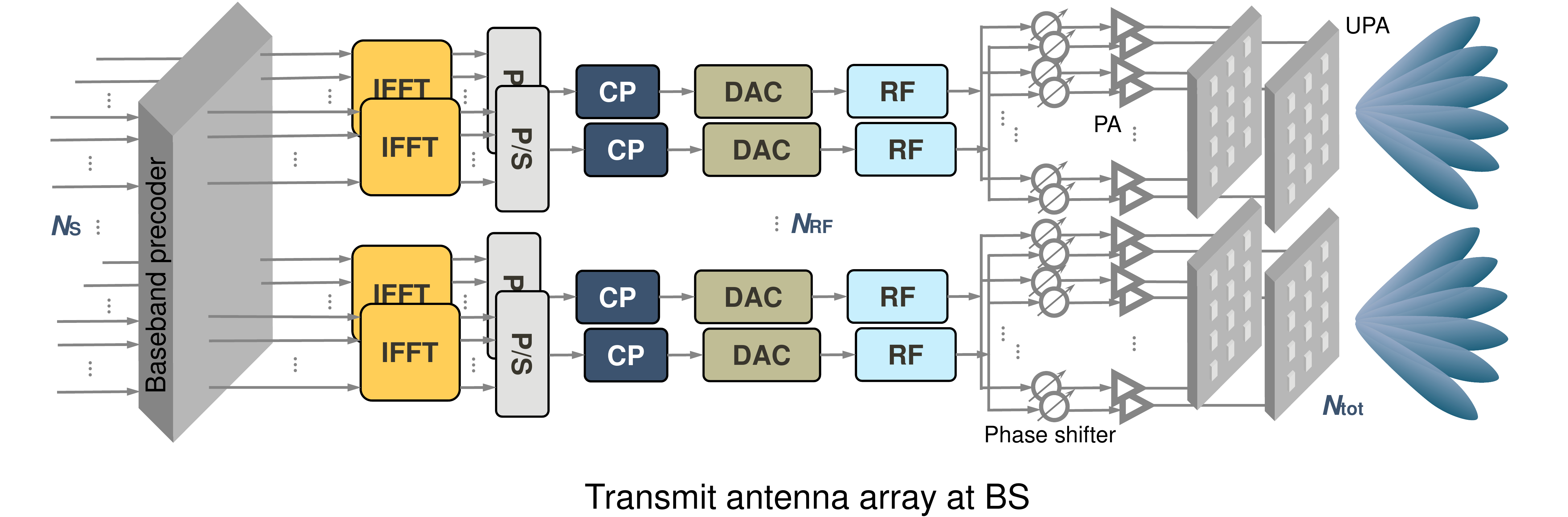}
\label{fig:subfigure1a}}
\quad
\subfigure[]{%
\includegraphics[width=4.05in]{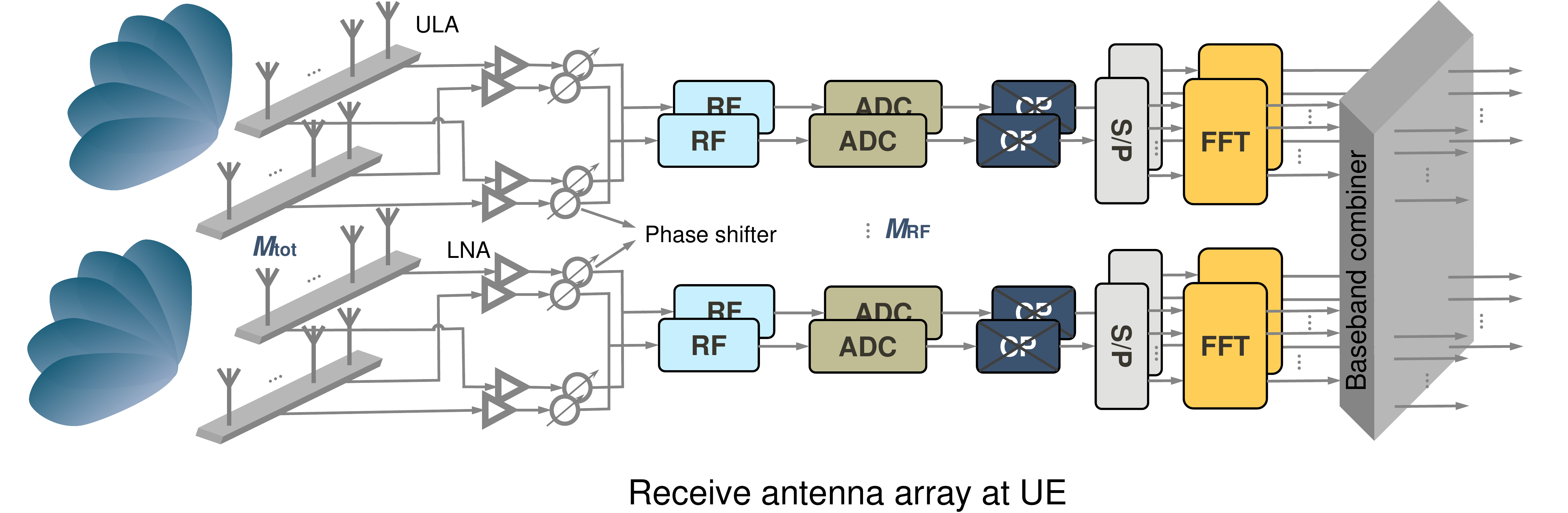}
\label{fig:subfigure1b}}
\caption{(a) Array-of-subarray architecture is employed at the BS with $N_{\mathrm{RF}}$ RF chains and $N_{\mathrm{tot}}$ transmit antenna elements. (b) Array-of-subarray architecture is employed at the UE with $M_{\mathrm{RF}}$ RF chains and $M_{\mathrm{tot}}$ receive antenna elements.}
\label{fig:figure1ab}
\end{figure}
\emph{1) Transceiver architecture, array configurations and synchronization signal structure.} We consider a precoded MIMO-OFDM system with $N$ subcarriers and a hybrid precoding transceiver structure as shown in Figs.~\ref{fig:subfigure1a} and \ref{fig:subfigure1b}, in which the BS deploys $N_{\mathrm{tot}}$ transmit antennas and $N_{\mathrm{RF}}$ RF chains, and the UE deploys $M_{\mathrm{tot}}$ receive antennas and $M_{\mathrm{RF}}$ RF chains. Both the BS and UE employ array-of-subarray architectures. As can be seen from Fig.~\ref{fig:figure1ab}, in an array-of-subarray architecture, a single RF chain controls an antenna subarray. Denote the number of antenna elements in each transmit subarray by $N_{\mathrm{A}}$ and the number of antenna elements in each receive subarray by $M_{\mathrm{A}}$. Then $N_{\mathrm{tot}}=N_{\mathrm{RF}}N_{\mathrm{A}}$ and $M_{\mathrm{tot}}=M_{\mathrm{RF}}M_{\mathrm{A}}$. For fully digital processing, $N_{\mathrm{A}}=M_{\mathrm{A}}=1$, while for single-stream analog-only processing, $N_{\mathrm{RF}}=M_{\mathrm{RF}}=1$.

Due to their constant amplitude and zero autocorrelation in both the time and frequency domains \cite{popov}, Zadoff-Chu (ZC)-type sequences are employed in this paper for the downlink synchronization signals. Denote the length of the employed ZC sequence by $N_{\mathrm{ZC}}$ and the sequence root index by $i$ ($i\in\left\{0,\cdots,N_{\mathrm{ZC}}-1\right\}$). For $m=0,\cdots,N_{\mathrm{ZC}}-1$, the sequence can be represented as
\begin{equation}\label{zcfdd}
s_{i}[m] = \exp\left\{-\mathrm{j}\frac{\pi m(m+1)i}{N_{\mathrm{ZC}}}\right\}.
\end{equation}
The cyclic auto-correlation of the ZC sequence results in a single dirac-impulse at zero-lag correlation, i.e.,
\begin{equation}\label{fcorrex}
\chi[\upsilon] = \sum_{m=0}^{N_{\mathrm{ZC}}-1}s_{i}[m]s_{i}^{*}[((m+\upsilon))_{N_{\mathrm{ZC}}}]=\delta[\upsilon],\hspace{4mm}\upsilon=0,\cdots,N_{\mathrm{ZC}}-1.
\end{equation}
The UE can therefore use this property to detect the correct frame timing position. In practice, the channel variations, noise power, and other impairments will affect the actual correlation values. Especially under low-resolution quantization, the good correlation properties of the ZC sequence are severely deteriorated by the quantization distortion. Besides the ZC-type sequences, Golay complementary sequences (GCSs) \cite{mgolay} exhibit a similar correlation property to (\ref{fcorrex}). As a Golay complementary pair contains two sequences, both cyclic prefix and cyclic postfix are needed in the GCSs to prevent the inter-symbol interference, while only the cyclic prefix is appended to the ZC sequence when it is propagated through the multi-path channels. Nevertheless, the GCSs can also be employed for downlink synchronization. The corresponding problem formulation, however, would be different from that in the ZC sequence design when exploiting the complementary sequence structure.

\begin{figure}
\centering
\includegraphics[width=3.45in]{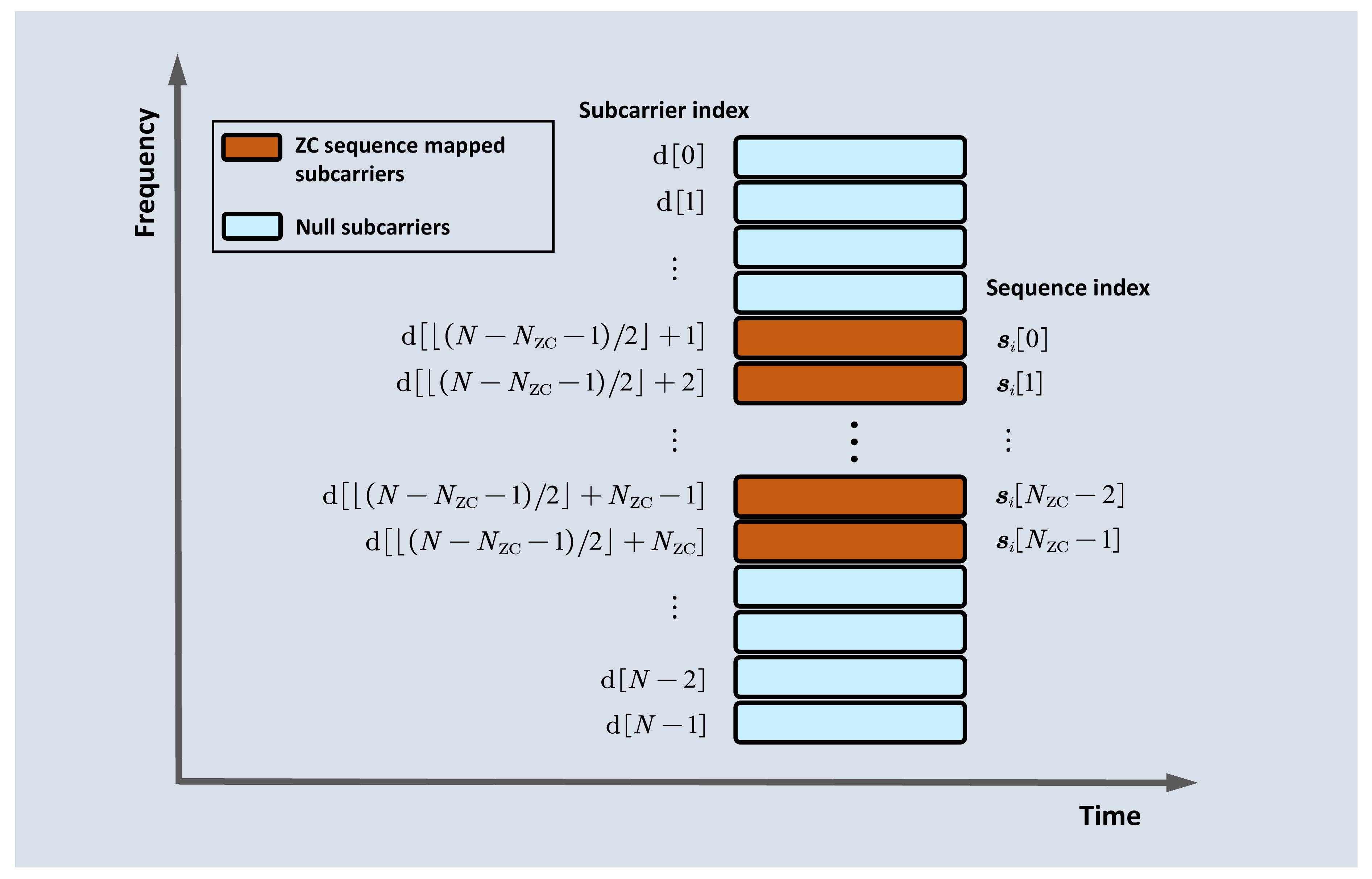}
\caption{A conceptual example of the mapping between the ZC sequence and the subcarriers in the frequency-domain. A length-$N_{\mathrm{ZC}}$ ZC sequence is mapped onto the central $N_{\mathrm{ZC}}$ subcarriers out of a total $N$ subcarriers.}
\label{fig:figure2}
\end{figure}
Denote the frequency-domain modulated symbol on subcarrier $k=0,\cdots,N-1$ by $\textsf{d}[k]$. We can then explicitly express the mapping between the ZC sequence and the subcarriers as
\begin{eqnarray}\label{eqq1}
\textsf{d}[\lfloor(N-N_{\mathrm{ZC}}-1)/2\rfloor+m+1]=\Bigg\{\begin{array}{l}
            s_{i}[m],\hspace{2mm} m=0,\cdots,N_{\mathrm{ZC}}-1,\\
            0,\hspace{2mm}\textrm{otherwise}.
          \end{array}
\end{eqnarray}
Note that (\ref{eqq1}) implies that in the frequency-domain, we map the ZC sequence onto the central $N_{\mathrm{ZC}}$ subcarriers (out of $N$ subcarriers) surrounding the DC-carrier symmetrically. In this paper, we set the DC-carrier as zero as in the LTE systems \cite{lte}; it is worth noting that no explicit DC-carrier is reserved for both the downlink and uplink in the 3GPP 5G NR systems (Release 15) \cite{5gnr}. We provide a conceptual example in Fig.~\ref{fig:figure2} to reveal this mapping relationship.

\begin{figure}
\centering
\includegraphics[width=5.25in]{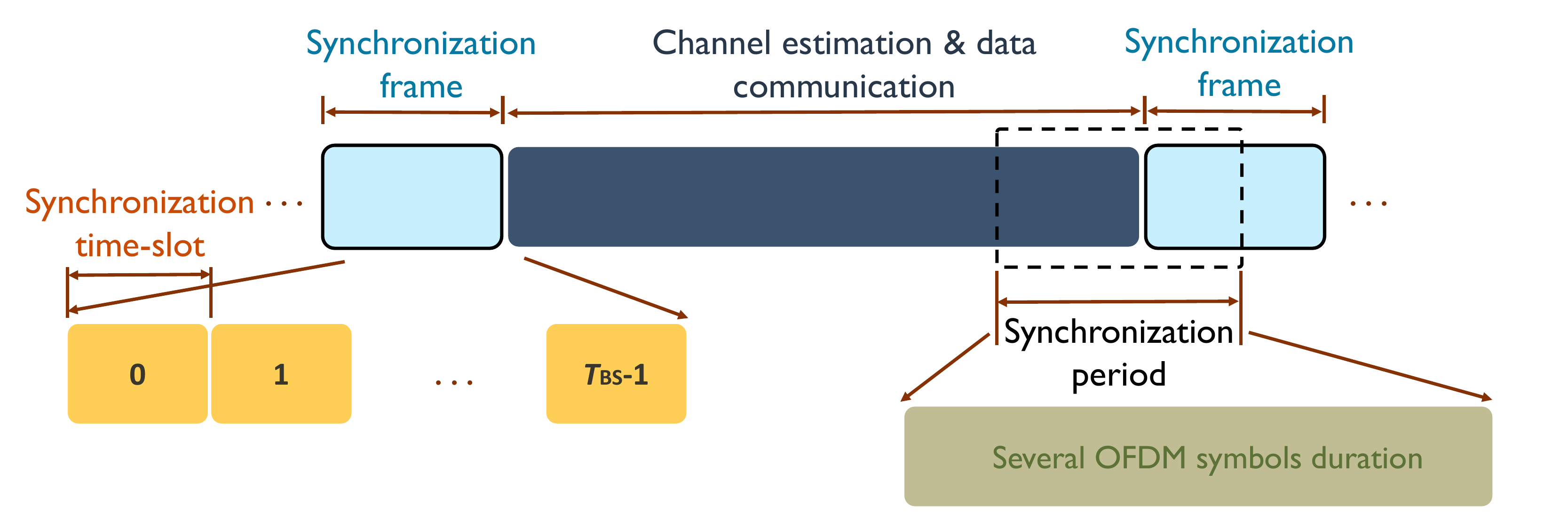}
\caption{A conceptual example of a potential synchronization frame structure. The BS-wise synchronization frame consists of $T_{\mathrm{BS}}$ synchronization time-slots. The length of one UE-wise synchronization period is equivalent to the duration of $T_{\mathrm{UE}}$ OFDM symbols.}
\label{fig:figure3}
\end{figure}
\emph{2) Synchronization frame structure and directional synchronization procedure}. Now, we explain the directional frame timing synchronization procedure. We start by introducing a potential frame structure, which is shown in Fig.~\ref{fig:figure3}. We define a time-slot, which may be one OFDM symbol duration ($T_{\mathrm{s}}$). We also define a synchronization frame, in which the BS transmits the downlink synchronization signals to the UEs. As can be seen from Fig.~\ref{fig:figure3}, each synchronization frame consists of $T_{\mathrm{BS}}$ synchronization time-slots. Different from uplink random access channels, the downlink multicast synchronization channels do not require guard intervals between the synchronization time-slots to deal with the propagation delay \cite{lte}. In conventional single-stream beamforming based approaches \cite{5gnr,5gbmgnt}, for a given synchronization time-slot, the BS probes one synchronization beam towards a predefined angular direction using a single RF chain. Across $T_{\mathrm{BS}}$ synchronization time-slots, the downlink synchronization beams fully scan the given angular range in a time-division multiplexing (TDM) manner.

Upon awakening from idle mode or power-up, the UE attempts to synchronize to the network and then performs a random access procedure. We therefore define a UE-wise synchronization period here, which is shown on the right-hand side in Fig.~\ref{fig:figure3}. For a given synchronization period, the UE employs fully digital front ends to detect the synchronization signal samples. In this paper, the length of one synchronization period is equivalent to the duration of $T_{\mathrm{UE}}$ OFDM symbols.
%In Fig.~\ref{fig:figure3}, we depict an overlapping region between the UE-wise synchronization period and the BS-wise synchronization frame. In our model, we assume that the overlapping region contains at least one BS-wise synchronization time-slot, which corresponds to a complete synchronization sequence. With this assumption, the UE can fully exploit the inherent correlation properties of the synchronization signal samples to perform the frame timing synchronization.
%The UE can only detect the synchronization signals probed during the overlapping region. By exploiting the inherent correlation properties of the synchronization signals, the UE then performs the frame timing synchronization.

%\begin{figure}
%\centering
%%\includegraphics[width=4.5in]{probing_frame_slot.pdf}
%\includegraphics[width=4.25in]{cexample1.pdf}
%\caption{For a given synchronization time-slot $0$, a BS employs an analog beamformer $\bm{f}_{0}$ to transmit the synchronization signal samples ($\left\{\textsf{d}[0],\cdots,\textsf{d}[N-1]\right\}$ in the frequency-domain) to a given UE $u$ ($u\in\left\{1,\cdots,N_{\mathrm{UE}}\right\}$). UE $u$ uses fully digital combining to receive the synchronization signal samples. For the $b$-th receive antenna at UE $u$, the received synchronization signal samples are $\left\{q^{(0)}_{u,b}[\mathrm{t}],\cdots,q^{(0)}_{u,b}[\mathrm{t}+N-1]\right\}$ among all the received signal samples $\left\{q^{(0)}_{u,b}[0],\cdots,q^{(0)}_{u,b}[NT_{\mathrm{UE}}-1]\right\}$.}
%\label{fig:figure4}
%\end{figure}
\emph{3) Received synchronization signal model}. Based on the employed array configurations and synchronization signal structure, we develop the received synchronization signal model assuming $N_{\mathrm{RF}}=1$, i.e., single-stream analog-only beamforming at the BS and $M_{\mathrm{RF}}=M_{\mathrm{tot}}$, i.e., fully digital baseband combining at the UE. Note that in Section IV, we will modify the received synchronization signal model by assuming $N_{\mathrm{RF}}>1$. We consider a given UE $u\in\left\{1,\cdots,N_{\mathrm{UE}}\right\}$ in a single cell, where $N_{\mathrm{UE}}$ corresponds to the total number of UEs in the cell of interest. For better illustration of the synchronization procedure, we assume $T_{\mathrm{BS}}=1$, i.e., a single synchronization time-slot, say, synchronization time-slot $0$. We further assume that all the synchronization signal samples $\left\{\textsf{d}[0],\cdots,\textsf{d}[N-1]\right\}$ probed during synchronization time-slot $0$ are received by the UE across the synchronization period. Based on these assumptions, we now derive the received signal model for our system.

The symbol vector $\textbf{\textsf{d}}$ in (\ref{eqq1}) is first transformed to the time-domain via $N$-point IFFTs, generating the discrete-time signals at symbol durations $n=0,\cdots,N-1$ as
\begin{equation}\label{tdzc}
d[n]=\frac{1}{\sqrt{N}}\sum_{k=0}^{N-1}\textsf{d}[k]e^{\mathrm{j}\frac{2\pi k}{N}n}.
\end{equation}
Before applying an $N_{\mathrm{tot}}\times 1$ wideband analog beamforming vector, a cyclic prefix (CP) is added to the symbol vector such that the length of the CP is greater than or equal to the maximum delay spread of the multi-path channels. Each sample in the symbol vector is then transmitted by a common wideband analog beamforming vector $\bm{f}_{0}$ probed from the BS, satisfying the power constraint $\left[\bm{f}_{0}\bm{f}_{0}^{*}\right]_{a,a}=\frac{1}{N_{\mathrm{tot}}}$, where $a=1,\cdots,N_{\mathrm{tot}}$. In this paper, we use superscript $(0)$ to denote variables obtained assuming $\bm{f}_{0}$.

Considering the $b$-th receive antenna ($b\in\left\{1,\cdots,M_{\mathrm{tot}}\right\})$ at UE $u$, we denote the time-domain received signal samples by $\bm{q}^{(0)}_{u,b}=\big[q^{(0)}_{u,b}[0],\cdots,q^{(0)}_{u,b}[NT_{\mathrm{UE}}-1]\big]^{\mathrm{T}}$. Note that we ignore the CPs here because they do not affect our proposed frame timing synchronization strategy. The CPs will be discarded after the timing synchronization to mitigate the inter-symbol interference, and the remaining samples are further processed for other initial access tasks. Denote the number of channel taps by $L_u$, the corresponding channel impulse response at tap $\ell\in\left\{0,\cdots,L_{u}-1\right\}$ by $\bm{H}_u[\ell]\in\mathbb{C}^{M_{\mathrm{tot}}\times N_{\mathrm{tot}}}$, and the additive white Gaussian noise by $w_u[n]\sim \mathcal{N}_{c}(0,\sigma^{2})$. We incorporate the effect of carrier frequency offset (CFO) in the received signal model. The CFO is a result of frequency mismatch between the transceiver's oscillators and the Doppler shift. Denote the frequency mismatch with respect to the subcarrier spacing by $\varepsilon_u$. As the UE employs fully digital baseband processing, each receive antenna first quantizes the received synchronization signals with dedicated ADCs. Denote $\mathcal{Q}(\cdot)$ as the quantization function. Further, denote the index of the first synchronization signal sample in the received signal by $\mathrm{t}\in\left\{0,\cdots,N(T_{\mathrm{UE}}-1)\right\}$. For $n=0,\cdots,N-1$, the received samples are
\begin{eqnarray}\label{zcmap16}
q^{(0)}_{u,b}[\mathrm{t}+n]=\mathcal{Q}\left(e^{\mathrm{j}\frac{2\pi\varepsilon_u}{N}n}\sum_{\ell=0}^{L_u-1}\left[\bm{H}_u[\ell]\right]_{b,:}\bm{f}_{0}d[((n-\ell))_{N}]+w_u[n]\right).
\end{eqnarray}
Practical cellular networks such as the LTE systems \cite{lte} perform the frame timing synchronization in the presence of CFO. After correctly detecting the frame timing, the UE then estimates the CFO and conducts the frequency synchronization. In this paper, we set the offset $\varepsilon_u=0$ ($u=1,\cdots,N_{\mathrm{UE}}$) in the following derivations. This is mainly because the CFO does not affect the development of the proposed frame timing synchronization algorithm in Section IV, though it affects the overall frame timing synchronization performance. In Section V, we provide a simulation plot to characterize the CFO effect on the timing synchronization performance. Neglecting the CFO, we rewrite (\ref{zcmap16}) as
\begin{eqnarray}\label{zcmap17}
q^{(0)}_{u,b}[\mathrm{t}+n]=\mathcal{Q}\Bigg(\underbrace{\sum_{\ell=0}^{L_u-1}\left[\bm{H}_u[\ell]\right]_{b,:}\bm{f}_{0}d[((n-\ell))_{N}]+w_u[n]}_{y^{(0)}_{u,b}[\mathrm{t}+n]}\Bigg).
\end{eqnarray}
The received signal samples $\{q^{(0)}_{u,b}[0],\cdots,q^{(0)}_{u,b}[NT_{\mathrm{UE}}-1]\}$ also contain non-synchronization signal samples. Hence, we have
\begin{equation}\label{smasum}
q^{(0)}_{u,b}[m']=\Bigg\{\begin{array}{l}
                                    \mathcal{Q}\left(\sum_{\ell=0}^{L_u-1}\left[\bm{H}_u[\ell]\right]_{b,:}\bm{f}_{0}d[((m'-\mathrm{t}-\ell))_{N}]+w_u[m'-\mathrm{t}]\right), m' = \mathrm{t},\cdots,\mathrm{t}+N-1, \\
\mathcal{Q}(w_u[m']), \hspace{2mm}\textrm{otherwise}.
                                  \end{array}
\end{equation}
As can be seen from (\ref{smasum}), we model the received non-synchronization signal samples as noise, though they may contain deterministic data and/or control information.
%Note that in practical systems, the periodicity of the BS-wise synchronization frame is much greater than the UE-wise synchronization period \cite{lte}. That is, for a given UE-wise synchronization period, the UE can only receive the synchronization signals probed during one BS-wise synchronization frame.

By using the discrete-time received signal vector $\bm{q}^{(0)}_{u,b}$ and the known unquantized reference synchronization sequence, the frame timing synchronization can be conducted by UE $u$. According to (\ref{tdzc}), the unquantized reference synchronization sequence locally stored at the UE is $\bm{d} = \left[d[0],d[1],\cdots,d[N-1]\right]^{\mathrm{T}}$. The UE then calculates the time-domain cross-correlation between the received signal samples and the unquantized reference synchronization sequence for the $b$-th receive antenna as
\begin{equation}\label{crosscorr}
\Gamma^{(0)}_{u,b}[\nu] = \sum_{n=0}^{N-1}q^{(0)}_{u,b}[n+\nu]d^{*}[n],
\end{equation}
where $\nu=0,\cdots,N(T_{\mathrm{UE}}-1)$. Denote the index of the selected receive antenna by $\hat{b}$. The maximum likelihood detector \cite{tz07} finds the estimate of the frame timing position $\hat{\nu}$ that corresponds to the peak in the correlation, i.e.,
\begin{equation}\label{corropt}
\left(\hat{\nu},\hat{b}\right)=\underset{\substack{\nu=0,\cdots,N(T_{\mathrm{UE}}-1)\\ b=1,\cdots,M_{\mathrm{tot}}}}{\mathrm{argmax}}\left|\Gamma^{(0)}_{u,b}[\nu]\right|^2.
\end{equation}
If $\mathcal{Q}(\cdot)$ in (\ref{zcmap17}) and (\ref{smasum}) corresponds to low-resolution quantization (e.g., $1$-$4$ bits), the corresponding quantization distortion will damage the good correlation properties of the employed synchronization sequences, leading to degraded timing synchronization performance.
\subsection{Channel model}
Assume that the channel between the BS and UE $u\in\left\{1,\cdots,N_{\mathrm{UE}}\right\}$ has $R_u$ paths, and each path $r$ has azimuth and elevation angle-of-departures (AoDs) $\phi_{u,r}$, $\theta_{u,r}$ and angle-of-arrival (AoA) $\psi_{u,r}$. Let $p(\tau)$ denote the combined effect of filtering and pulse shaping for $T_{\mathrm{s}}$-spaced signaling at $\tau$ seconds. We then express the time-domain delay-$\ell$ MIMO channel matrix as
\begin{eqnarray}\label{delayd}
\bm{H}_u[\ell]=\sum_{r=1}^{R_u}\beta_{u,r}p\left(\ell T_{\mathrm{s}}-\tau_{r}\right)\bm{a}_{\mathrm{rx}}(\psi_{u,r})\bm{a}_{\mathrm{tx}}^{*}(\theta_{u,r},\phi_{u,r}),
\end{eqnarray}
where $\beta_{u,r}$ represents the complex path gain of path-$r$ between the BS and UE $u$, and $\bm{a}_{\mathrm{rx}}(\cdot)\in\mathbb{C}^{M_{\mathrm{tot}}\times1}$ and $\bm{a}_{\mathrm{tx}}(\cdot,\cdot)\in\mathbb{C}^{N_{\mathrm{tot}}\times1}$ correspond to the receive and transmit array response vectors. For instance, if the BS employs a uniform planar array (UPA) in the $\mathrm{xy}$-plane and the UE uses a uniform linear array (ULA) on the $\mathrm{y}$ axis, $\bm{a}_{\mathrm{tx}}(\cdot,\cdot)$ and $\bm{a}_{\mathrm{rx}}(\cdot)$ would exhibit the same structures as those in (3) and (4) in \cite{dztrans2d}. Note that the proposed design approach does not depend on array geometry. Define $\bm{A}^{\mathrm{RX}}_{u}$ and $\bm{A}^{\mathrm{TX}}_{u}$ as
\begin{eqnarray}
&\bm{A}^{\mathrm{RX}}_{u}=\left[\bm{a}_{\mathrm{rx}}(\psi_{u,1})\hspace{2mm}\bm{a}_{\mathrm{rx}}(\psi_{u,2})\hspace{2mm}\cdots\hspace{2mm}\bm{a}_{\mathrm{rx}}(\psi_{u,R_u})\right],&\\
&\bm{A}^{\mathrm{TX}}_{u}=\left[\bm{a}_{\mathrm{tx}}(\theta_{u,1},\phi_{u,1})\hspace{2mm}\bm{a}_{\mathrm{tx}}(\theta_{u,2},\phi_{u,2})\hspace{2mm}\cdots\hspace{2mm}\bm{a}_{\mathrm{tx}}(\theta_{u,R_u},\phi_{u,R_u})\right],&
\end{eqnarray}
which contain the receive and transmit array response vectors and a diagonal matrix $\bm{G}_u[\ell]=\mathrm{diag}\left(\left[g_{u,1,\ell},\cdots,g_{u,R_u,\ell}\right]^{\mathrm{T}}\right)$, where $g_{u,r,\ell}=\beta_{u,r}p\left(\ell T_{\mathrm{s}}-\tau_{r}\right)$ for $r\in\left\{1,\cdots,R_u\right\}$. We can then rewrite the channel matrix in (\ref{delayd}) in a more compact form as $\bm{H}_u[\ell]=\bm{A}^{\mathrm{RX}}_{u}\bm{G}_u[\ell]\left(\bm{A}^{\mathrm{TX}}_{u}\right)^{*}$. Denote the corresponding channel frequency response on subcarrier $k=0,\cdots,N-1$ by $\textbf{\textsf{G}}_u[k]$. We have $\textbf{\textsf{G}}_u[k] = \sum_{\ell=0}^{L_u-1}\bm{G}_u[\ell]e^{-\mathrm{j}2\pi\ell k/N}$ and $\textsf{g}_{u,r,k}=\left[\textbf{\textsf{G}}_u[k]\right]_{r,r}=\sum_{\ell=0}^{L_u-1}g_{u,r,\ell}e^{-\mathrm{j}2\pi\ell k/N}$. We can then express the corresponding frequency-domain channel matrix as $\textbf{\textsf{H}}_u[k]=\bm{A}^{\mathrm{RX}}_{u}\textbf{\textsf{G}}_u[k]\left(\bm{A}^{\mathrm{TX}}_{u}\right)^{*}$. In Sections III and IV, we use $\bm{H}_{u}[\ell]$ to develop the received synchronization signal model in the time-domain and $\textbf{\textsf{H}}_u[k]$ to illustrate the proposed algorithm in the frequency-domain.

\section{Optimization Problem Formulation for Directional Frame Timing Synchronization under Low-Resolution ADCs}

In this section, we first formulate the directional frame timing synchronization problem for mmWave systems operating with low-resolution ADCs. We then show that the formulated problem is a max-min multicast optimization problem, which cannot be effectively solved under the framework of single-stream beamforming. For clarity, we conduct the problem formulation using the frequency-domain representations, though we first present the received signal model in the time-domain.

\subsection{Optimization metric for low-resolution timing synchronization}
To formulate the optimization problem for low-resolution timing synchronization, we need to first determine a proper optimization metric. Similar to Section II-A, we assume a single synchronization time-slot, single-stream analog-only beamforming at the BS and fully digital baseband processing at the UE. By Bussgang's theorem \cite{buss1,buss2}, the quantization output in (\ref{zcmap17}) can be decoupled into a signal part and an uncorrelated distortion component. This decomposition is accurate in low and medium SNR ranges \cite{nyucell}. We first define $\bm{E}^{(0)}_{u,b}=\mathrm{diag}\Big(\Big[\eta^{(0)}_{u,b}[0],\cdots,\eta^{(0)}_{u,b}[N-1]\Big]^{\mathrm{T}}\Big)$ as the quantization distortion matrix with
\begin{equation}\label{defnmse}
\eta^{(0)}_{u,b}[n] = \frac{\mathbb{E}\left[\left(q^{(0)}_{u,b}[\mathrm{t}+n]\right)^{*}y^{(0)}_{u,b}[\mathrm{t}+n]\right]}{\mathbb{E}\left[\left|y^{(0)}_{u,b}[\mathrm{t}+n]\right|^2\right]},
\end{equation}
as the distortion factor of the quantization, which depends on the quantizer design, the number of quantization bits and the distribution of the input samples to the quantizer \cite{cmollen}. Rewriting
\begin{eqnarray}
\bm{d}_{\ell}= \left[d[((-\ell))_{N}],d[((1-\ell))_{N}],\cdots,d[((N-1-\ell))_{N}]\right]^{\mathrm{T}},
\end{eqnarray}
and $\bm{q}^{(0)}_{u,b}=\left[q^{(0)}_{u,b}[\mathrm{t}],\cdots,q^{(0)}_{u,b}[\mathrm{t}+N-1]\right]^{\mathrm{T}}$, we then decompose (\ref{zcmap17}) as
\begin{eqnarray}\label{acceptpapr}
\bm{q}^{(0)}_{u,b}=\bm{E}^{(0)}_{u,b}\Bigg(\underbrace{\sum_{\ell=0}^{L_u-1}\left[\bm{H}_u[\ell]\right]_{b,:}\bm{f}_{0}\bm{d}_{\ell}}_{\bm{v}^{(0)}_{u,b}}+\bm{w}_u\Bigg) + \check{\bm{w}}^{(0)}_{u,b}.
\end{eqnarray}
Denote the quantization mean squared error by $\xi_u$ assuming Gaussian signaling with unit variance \cite{hyadc}. We further denote the covariance matrix of the noiseless unquantized received signal $\bm{v}^{(0)}_{u,b}$ in (\ref{acceptpapr}) by $\bm{R}_{\bm{v}^{(0)}_{u,b}}$ and the additive quantization noise vector by $\check{\bm{w}}^{(0)}_{u,b}=\left[\check{w}^{(0)}_{u,b}[0],\cdots,\check{w}^{(0)}_{u,b}[N-1]\right]^{\mathrm{T}}$. As shown in \cite{hyadc}, the quantization distortion matrix $\bm{E}^{(0)}_{u,b}$ can then be computed as
\begin{eqnarray}\label{newcovxyz}
\bm{E}^{(0)}_{u,b} = (1-\xi_u)\mathrm{diag}\left(\bm{R}_{\bm{v}^{(0)}_{u,b}}+\sigma^2\bm{I}_{N}\right)^{-\frac{1}{2}}.
\end{eqnarray}
Denoting the unquantized received signal power matrix for the $b$-th receive antenna at UE $u$ by $\bm{V}^{(0)}_{u,b}$, we can express the covariance matrix of the quantization noise vector $\check{\bm{w}}^{(0)}_{u,b}$ as \cite{hyadc,jhhbadc}
\begin{eqnarray}\label{covvv}
\bm{R}_{\check{\bm{w}}^{(0)}_{u,b}}=\bm{E}^{(0)}_{u,b}\left(\bm{I}_{N}-\bm{E}^{(0)}_{u,b}\right)\underbrace{\mathrm{diag}\left(\bm{R}_{\bm{v}^{(0)}_{u,b}}+\sigma^2\bm{I}_{N}\right)}_{\bm{V}^{(0)}_{u,b}}.
\end{eqnarray}
As can be seen from (\ref{newcovxyz}) and (\ref{covvv}), both $\bm{E}^{(0)}_{u,b}$ and $\bm{R}_{\check{\bm{w}}^{(0)}_{u,b}}$ depend on $\bm{R}_{\bm{v}^{(0)}_{u,b}}$, which depends on the effective beam-space channel.
%The values of the quantization NMSE $\xi_u$ for various numbers of quantization bits are listed in \cite[Table~I]{chestjm}.
%With low-resolution ADCs, the good correlation properties of the employed synchronization signals would be contaminated by the quantization distortion. This would reduce the corresponding correlation peak resulting in increased probability of miss detection of the frame timing position. Hence, by increasing the received synchronization SQNR at zero-lag correlation, we expect to improve the overall synchronization performance under low-resolution ADCs.

In the following, we compute the zero-lag correlation between the received signal samples and the known unquantized reference synchronization sequence in the frequency-domain. We first express the frequency-domain quantized received signal $\textbf{\textsf{q}}^{(0)}_{u,b}=\left[\textsf{q}^{(0)}_{u,b}[0],\cdots,\textsf{q}^{(0)}_{u,b}[N-1]\right]^{\mathrm{T}}$ as
\begin{eqnarray}
\textsf{q}^{(0)}_{u,b}[k]&=&\eta^{(0)}_{u,b}[k]\left[\bm{A}^{\mathrm{RX}}_{u}\textbf{\textsf{G}}_u[k]\left(\bm{A}^{\mathrm{TX}}_{u}\right)^{*}\right]_{b,:}\bm{f}_{0}\textsf{d}[k]+\eta^{(0)}_{u,b}[k]\textsf{w}_u[k]+\check{\textsf{w}}^{(0)}_{u,b}[k],
\end{eqnarray}
where $\check{\textsf{w}}^{(0)}_{u,b}[k]=\sum_{n=0}^{N-1}\check{w}^{(0)}_{u,b}[n]e^{-\mathrm{j}2\pi nk/N}$. We then calculate the zero-lag frequency-domain correlation between $\textbf{\textsf{q}}^{(0)}_{u,b}$ and the unquantized reference synchronization sequence $\textbf{\textsf{d}}$ as
\begin{eqnarray}\label{zerolagfreq}
\Lambda^{(0)}_{u,b}[0] &=& \sum_{k=0}^{N-1}\textsf{q}^{(0)}_{u,b}[k]\textsf{d}^{*}[k]\\
&=&\sum_{k=0}^{N-1}\eta^{(0)}_{u,b}[k]\left[\bm{A}^{\mathrm{RX}}_{u}\textbf{\textsf{G}}_u[k]\left(\bm{A}^{\mathrm{TX}}_{u}\right)^{*}\right]_{b,:}\bm{f}_{0}\textsf{d}[k]\textsf{d}^{*}[k]\nonumber\\
&+&\sum_{k=0}^{N-1}\eta^{(0)}_{u,b}[k]\textsf{w}_u[k]\textsf{d}^{*}[k]+\sum_{k=0}^{N-1}\check{\textsf{w}}^{(0)}_{u,b}[k]\textsf{d}^{*}[k].
\end{eqnarray}
Similar to (\ref{crosscorr}), we have $\hat{b}=\underset{b=1,\cdots,M_{\mathrm{tot}}}{\mathrm{argmax}}\left|\Lambda^{(0)}_{u,b}[0]\right|^2$.
%Note that we employ the receive antenna selection here due to the lack of channel knowledge during the initial synchronization phase, though other possible digital signal processing techniques can be exploited as well.
%Using the combining vector $\bm{c}_u$ in (\ref{zcmap17}) to interpret the receive antenna selection operation, we have $\left[\bm{c}_u\right]_{\hat{b}}=1$ and $\left[\bm{c}_u\right]_{b}=0$ for $b=1,\cdots,M_{\mathrm{tot}}$, $b\neq\hat{b}$.

Different from high-rate data communications, the synchronization signals usually occupy a relatively small portion of the entire bandwidth with continuous subcarriers surrounding the DC-carrier. For instance, in the LTE systems \cite{lte}, the synchronization signal samples occupy $62$ subcarriers (out of $1024$ for $10$ MHz bandwidth, or $2048$ for $20$ MHz bandwidth) surrounding the DC-carrier. If the same design principle applies to mmWave systems, along with the sparse nature of the mmWave channels, the synchronization signals will most likely experience ``flat'' channels instead of severe frequency selectivity. It is also worth noting that the ZC-type sequences are robust to the frequency selectivity \cite{zcfreq}.

Leveraging the flat synchronization channels assumption and denoting the frequency-domain counterpart of the unquantized received signal power matrix $\bm{V}^{(0)}_{u,\hat{b}}$ in (\ref{covvv}) by $\textbf{\textsf{V}}^{(0)}_{u,\hat{b}}$, we can first obtain
\begin{eqnarray}\label{dzzzzz}
\textbf{\textsf{V}}^{(0)}_{u,\hat{b}}&=&\textsf{g}_{u}^{2}\left|\left[\bm{a}_{\mathrm{rx}}(\psi_u)\bm{a}^{*}_{\mathrm{tx}}(\theta_u,\phi_u)\right]_{\hat{b},:}\bm{f}_{0}\right|^2\mathrm{diag}\left(\textbf{\textsf{d}}\textbf{\textsf{d}}^{*}\right)+\sigma^2\bm{I}_{N}\\
&=&\textsf{g}_{u}^{2}\left|\left[\bm{a}_{\mathrm{rx}}(\psi_u)\bm{a}^{*}_{\mathrm{tx}}(\theta_u,\phi_u)\right]_{\hat{b},:}\bm{f}_{0}\right|^2\nonumber\\ \label{dzzzzzz}
&\times&\mathrm{diag}\left(\left[\textsf{d}[0]\textsf{d}^{*}[0],\cdots,\textsf{d}[N-1]\textsf{d}^{*}[N-1]\right]^{\mathrm{T}}\right)+\sigma^2\bm{I}_{N},
\end{eqnarray}
where the path and subcarrier indices are dropped. Also because of the flat synchronization channels assumption, we define $\underline{\eta}^{(0)}_{u,\hat{b}}=\eta^{(0)}_{u,\hat{b}}[0]=\cdots=\eta^{(0)}_{u,\hat{b}}[N-1]$ as a common quantization distortion factor. By exploiting the inherent correlation properties of the ZC sequence design in (\ref{fcorrex}), we can compute the quantization noise power using (\ref{newcovxyz}), (\ref{covvv}) and (\ref{dzzzzzz}). We can then formulate the corresponding received synchronization SQNR at zero-lag correlation for UE $u$ as
\begin{eqnarray}\label{snqr0}
\gamma^{(0)}_{u,\hat{b}}=\frac{\underline{\eta}^{(0)}_{u,\hat{b}}\textsf{g}_{u}^{2}\left|\left[\bm{a}_{\mathrm{rx}}(\psi_u)\bm{a}^{*}_{\mathrm{tx}}(\theta_u,\phi_u)\right]_{\hat{b},:}\bm{f}_{0}\right|^2}{\underline{\eta}^{(0)}_{u,\hat{b}}\sigma^2+\left(1-\underline{\eta}^{(0)}_{u,\hat{b}}\right)\left(\textsf{g}_{u}^{2}\left|\left[\bm{a}_{\mathrm{rx}}(\psi_u)\bm{a}^{*}_{\mathrm{tx}}(\theta_u,\phi_u)\right]_{\hat{b},:}\bm{f}_{0}\right|^2+\sigma^2\right)}.
\end{eqnarray}

\begin{figure}
\centering
%\subfigure[]{%
\includegraphics[width=5.55in]{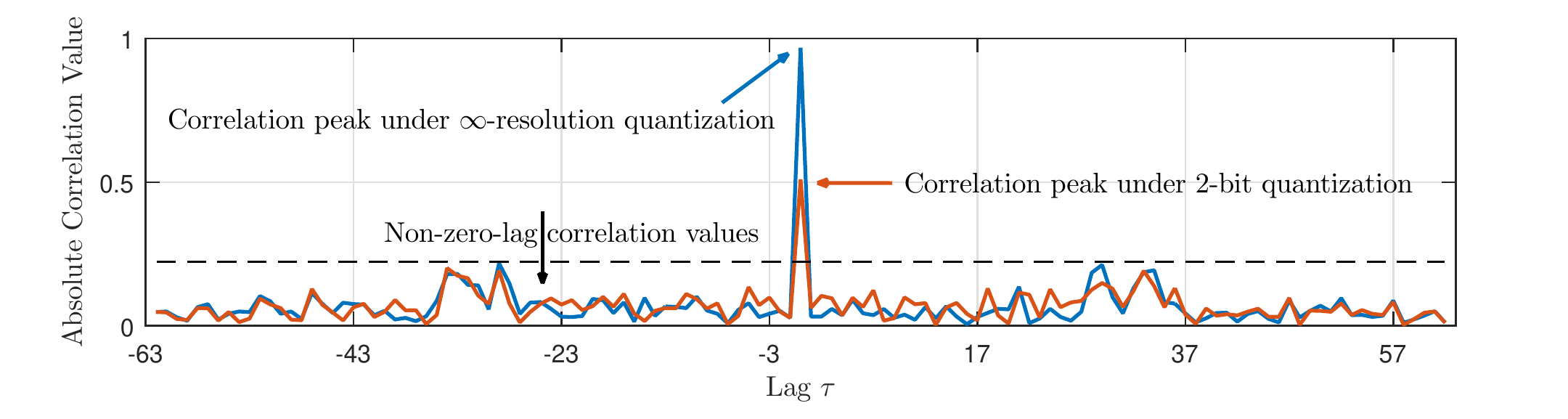}
\label{fig:subfigurePSSa}
%}
%\quad
%\subfigure[]{%
%\includegraphics[width=5.55in]{Fig_PSS_5.pdf}
%\label{fig:subfigurePSSb}}
\caption{Absolute correlation values of the employed synchronization sequence under both infinite-resolution and $2$-bit ADCs. To better characterize the impact of the quantization distortion on the absolute correlation values, a simple AWGN channel is considered with $0$ dB SNR. The transmit beamforming is not incorporated. A length-$62$ ZC sequence with root index $34$ is used.}
\label{fig:figurepss}
\end{figure}
Similar to the calculation of $\Lambda^{(0)}_{u,b}[0]$ in (\ref{zerolagfreq}), we can also compute the non-zero-lag correlation values. In Fig.~\ref{fig:figurepss}, we plot the absolute correlation values of the employed synchronization sequence under both infinite-resolution and $2$-bit ADCs. Both zero-lag and non-zero-lag correlations are revealed in this example. As can be seen from Fig.~\ref{fig:figurepss}, for both infinite-resolution and $2$-bit ADCs, the non-zero-lag correlation values have small magnitudes and exhibit similar patterns. For the zero-lag correlation, however, the correlation peak obtained under $2$-bit ADCs is much smaller than that under $\infty$-resolution quantization. Leveraging these observations, we employ the received synchronization SQNR at zero-lag correlation as the main optimization metric and formulate the corresponding optimization problems in Section III-B.

\subsection{Optimization problems for low-resolution timing synchronization}
We first consider a single UE, e.g., UE $u$ and a single synchronization time-slot, e.g., synchronization time-slot $0$. Our design target here is to maximize the received synchronization SQNR at zero-lag correlation for UE $u$. Note that in (\ref{snqr0}), the only parameter that can be tuned is the beamforming vector. To better compensate for the quantization distortion, it is therefore desirable to custom design the synchronization beams as long as necessary information is available at the BS. This is different from traditional TDM probing based approaches \cite{5gnr,5gbmgnt}, where the synchronization beams are predetermined and fixed. According to (\ref{snqr0}), we can formulate this maximization problem as
\begin{eqnarray}\label{optsingleue}
&\mathcal{P}0:\hspace{2mm}\underset{\bm{f}_{0}}{\mathrm{max}}\left\{\gamma^{(0)}_{u,\hat{b}}\right\}&\\
&\textrm{s.t.}\hspace{2mm}\left[\bm{f}_{0}\bm{f}_{0}^{*}\right]_{a,a}=\frac{1}{N_{\mathrm{tot}}}, a=1,\cdots,N_{\mathrm{tot}}.&\nonumber
\end{eqnarray}
To simplify $\mathcal{P}0$, we assume that the BS uses a predefined analog beam codebook $\mathcal{F}$. We can then reformulate (\ref{optsingleue}) as
\begin{eqnarray}\label{optsingleuesimp}
&\mathcal{P}1:\hspace{2mm}\underset{\bm{f}_{0}}{\mathrm{max}}\left\{\gamma^{(0)}_{u,\hat{b}}\right\}&\\
&\textrm{s.t.}\hspace{2mm}\bm{f}_{0}\in\mathcal{F}.&\nonumber
\end{eqnarray}
Solving $\mathcal{P}1$ does not increase the non-zero-lag correlation values at the same pace as the zero-lag peak correlation value. We verify this by the following lemma, which shows that maximizing the received synchronization SQNR at zero-lag correlation also maximizes the power difference between the zero-lag peak correlation value and the non-zero-lag correlation values assuming low-resolution quantization.
\begin{lemma1}
Consider the $\hat{b}$-th receive antenna at UE $u$ equipped with low-resolution ADCs. Assume $\bm{f}_0$ and denote the corresponding power ratio between the zero-lag correlation value and a non-zero-lag correlation value by $\varsigma^{(0)}_{u,\hat{b}}$. For $\bm{f}^{\star}_{0}$, if the resulted received synchronization SQNR at zero-lag correlation $\gamma^{(0)^{\star}}_{u,\hat{b}}=\mathrm{max}\left\{\gamma^{(0)}_{u,\hat{b}},\bm{f}_{0}\in\mathcal{F}\right\}$, then $\varsigma^{(0)^{\star}}_{u,\hat{b}}=\mathrm{max}\left\{\varsigma^{(0)}_{u,\hat{b}},\bm{f}_{0}\in\mathcal{F}\right\}$.
\end{lemma1}
\begin{proof}
See Appendix.
\end{proof}
Next, we extend the problem formulation to a single cell with multiple UEs. In this case, we expect that for a given synchronization time-slot, a group of UEs can simultaneously synchronize to the network with satisfying synchronization performance. This reduces the overall access delay of the network. The design target therefore becomes maximizing the minimum received synchronization SQNR at zero-lag correlation for all potential UEs. Assuming a total of $N_{\mathrm{UE}}$ UEs and synchronization time-slot $0$, we formulate the following max-min optimization problem as
\begin{eqnarray}\label{realmaxminop}
&\mathcal{P}2:\hspace{2mm}\underset{\bm{f}_{0}}{\mathrm{max}}\hspace{2mm}\underset{\forall u}{\mathrm{min}}\left\{\gamma^{(0)}_{u,\hat{b}}\right\}&\\
&\textrm{s.t.}\hspace{2mm}\bm{f}_{0}\in\mathcal{F}.&\nonumber
%&\textrm{s.t.}\hspace{2mm}\left[\bm{f}_{0}\left(\bm{f}_{0}\right)^{*}\right]_{a,a}=\frac{1}{N_{\mathrm{tot}}}, a=1,\cdots,N_{\mathrm{tot}}.&\nonumber
\end{eqnarray}
Denoting by $\lambda_{u}=\sigma^2/\textsf{g}_{u}^{2}\left|\left[\bm{a}_{\mathrm{rx}}(\psi_u)\right]_{\hat{b}}\right|^2$, we rewrite $\gamma^{(0)}_{u,\hat{b}}$ in (\ref{snqr0}) as
\begin{eqnarray}
\gamma^{(0)}_{u,\hat{b}}&=&\frac{\underline{\eta}^{(0)}_{u,\hat{b}}\left|\bm{a}^{*}_{\mathrm{tx}}(\theta_u,\phi_u)\bm{f}_{0}\right|^2}{\underline{\eta}^{(0)}_{u,\hat{b}}\frac{\sigma^2}{\textsf{g}_{u}^{2}\left|\left[\bm{a}_{\mathrm{rx}}(\psi_u)\right]_{\hat{b}}\right|^2}+\left(1-\underline{\eta}^{(0)}_{u,\hat{b}}\right)\left(\left|\bm{a}^{*}_{\mathrm{tx}}(\theta_u,\phi_u)\bm{f}_{0}\right|^2+\frac{\sigma^2}{\textsf{g}_{u}^{2}\left|\left[\bm{a}_{\mathrm{rx}}(\psi_u)\right]_{\hat{b}}\right|^2}\right)}\\
&=&\frac{\underline{\eta}^{(0)}_{u,\hat{b}}\left|\bm{a}^{*}_{\mathrm{tx}}(\theta_u,\phi_u)\bm{f}_{0}\right|^2}{\underline{\eta}^{(0)}_{u,\hat{b}}\lambda_u+\left(1-\underline{\eta}^{(0)}_{u,\hat{b}}\right)\left(\left|\bm{a}^{*}_{\mathrm{tx}}(\theta_u,\phi_u)\bm{f}_{0}\right|^2+\lambda_u\right)}.\label{qabbc}
\end{eqnarray}
We can interpret $\lambda_{u}$ as the inverse of the received SNR at UE $u$. Denoting by $\lambda_{\mathrm{max}}=\mathrm{max}\big\{\lambda_1,\cdots,\\ \lambda_{N_{\mathrm{UE}}}\big\}$ and replacing $\lambda_u$ in (\ref{qabbc}) with $\lambda_{\mathrm{max}}$, we define a lower bound of $\gamma^{(0)}_{u,\hat{b}}$ as
\begin{equation}\label{maxmin0}
\breve{\gamma}^{(0)}_{u,\hat{b}}=\frac{\underline{\eta}^{(0)}_{u,\hat{b}}\left|\bm{a}^{*}_{\mathrm{tx}}(\theta_u,\phi_u)\bm{f}_{0}\right|^2}{\underline{\eta}^{(0)}_{u,\hat{b}}\lambda_{\mathrm{max}}+\left(1-\underline{\eta}^{(0)}_{u,\hat{b}}\right)\left(\left|\bm{a}^{*}_{\mathrm{tx}}(\theta_u,\phi_u)\bm{f}_{0}\right|^2+\lambda_{\mathrm{max}}\right)}.
\end{equation}
For any given UE $u$, we therefore have
\begin{equation}\label{lessbound0}
\breve{\gamma}^{(0)}_{u,\hat{b}}\leq\gamma^{(0)}_{u,\hat{b}}.
\end{equation}
By plugging the results of (\ref{newcovxyz}) and (\ref{dzzzzzz}) into (\ref{maxmin0}), we obtain
\begin{equation}\label{lubound0}
\breve{\gamma}^{(0)}_{u}=\frac{\left|\bm{a}^{*}_{\mathrm{tx}}(\theta_u,\phi_u)\bm{f}_{0}\right|^2}{\lambda_{\mathrm{max}}+\left[\frac{\left[\sigma^2\left(\left|\bm{a}^{*}_{\mathrm{tx}}(\theta_u,\phi_u)\bm{f}_{0}\right|^{2}/\lambda_{\mathrm{max}}+1\right)\right]^{1/2}}{1-\xi_u}-1\right]\left(\left|\bm{a}^{*}_{\mathrm{tx}}(\theta_u,\phi_u)\bm{f}_{0}\right|^2+\lambda_{\mathrm{max}}\right)},
\end{equation}
which becomes irrelevant to the selected receive antenna index $\hat{b}$ for UE $u$. Denoting by $\xi_{\mathrm{max}}=\left\{\xi_1,\cdots,\xi_{N_{\mathrm{UE}}}\right\}$ and replacing $\xi_u$ in (\ref{lubound0}) with $\xi_{\mathrm{max}}$, we can further define a lower bound of $\breve{\gamma}_{u}^{(0)}$ as
\begin{equation}\label{maxminone}
\acute{\gamma}^{(0)}_{u}=\frac{\left|\bm{a}^{*}_{\mathrm{tx}}(\theta_u,\phi_u)\bm{f}_{0}\right|^2}{\lambda_{\mathrm{max}}+\left[\frac{\left[\sigma^2\left(\left|\bm{a}^{*}_{\mathrm{tx}}(\theta_u,\phi_u)\bm{f}_{0}\right|^{2}/\lambda_{\mathrm{max}}+1\right)\right]^{1/2}}{1-\xi_{\mathrm{max}}}-1\right]\left(\left|\bm{a}^{*}_{\mathrm{tx}}(\theta_u,\phi_u)\bm{f}_{0}\right|^2+\lambda_{\mathrm{max}}\right)}.
\end{equation}
That is, for any given UE $u$, we have
\begin{equation}\label{lessbound1}
\acute{\gamma}^{(0)}_{u}\leq\breve{\gamma}^{(0)}_{u}.
\end{equation}
Note that $1/\lambda_{\mathrm{max}}$ and $1-\xi_{\mathrm{max}}$ represent the lowest received SNR and the lowest quantization resolution among all UEs, and they can be used to characterize the worst-case scenario of the network. Based on (\ref{lessbound0}) and (\ref{lessbound1}), we can therefore reformulate the optimization problem in (\ref{realmaxminop}) as
\begin{eqnarray}\label{realmaxminoprev}
&\mathcal{P}3:\hspace{2mm}\underset{\bm{f}_{0}}{\mathrm{max}}\hspace{2mm}\underset{\forall u}{\mathrm{min}}\left\{\acute{\gamma}^{(0)}_{u}\right\}&\\
&\textrm{s.t.}\hspace{2mm}\bm{f}_{0}\in\mathcal{F}.&\nonumber
%&\textrm{s.t.}\hspace{2mm}\left[\bm{f}_{0}\bm{f}_{0}^{*}\right]_{a,a}=\frac{1}{N_{\mathrm{tot}}}, a=1,\cdots,N_{\mathrm{tot}}.&\nonumber
\end{eqnarray}
%\subsection{Asymptotic solution assuming channel knowledge}
Solving (\ref{realmaxminoprev}) requires explicit knowledge of $\theta_{u}$'s, $\phi_u$'s, $\lambda_{u}$'s ($\lambda_{\mathrm{max}}$), and $\xi_{u}$'s ($\xi_{\mathrm{max}}$) for all UEs ($u=1,\cdots,N_{\mathrm{UE}}$). In practice, $\lambda_{\mathrm{max}}$ and $\xi_{\mathrm{max}}$ can be replaced with predefined system-specific values that characterize the worst-case scenario of the network. The explicit channel directional information, however, is unavailable during the initial timing synchronization phase. In addition, a practical beam codebook $\mathcal{F}$ usually has a limited number of candidate beam codewords. Hence, for conventional single-stream multicast beamforming, the corresponding synchronization performance is highly limited by the beam codebook resolution especially under low-resolution quantization.
%To improve the synchronization performance under low-resolution ADCs, we need to design new beamforming strategies and synchronization signal structures to better trade-off the effective beam-space channel gain and the quantization distortion. For single-stream analog-only beamforming, the room for optimizing the beamforming strategy to further improve the synchronization performance under low-resolution quantization is small.
%\section{Proposed multi-beam probing with common synchronization signal design}
\section{Proposed Directional Frame Timing Synchronization Design under Low-Resolution ADCs}
%\begin{figure}
%\centering
%%\includegraphics[width=4.5in]{probing_frame_slot.pdf}
%\includegraphics[width=4.35in]{BS_UE_relation_multi.pdf}
%\caption{A conceptual example of the proposed multi-beam probing based directional synchronization design. A single cell is considered with a total of $N_{\mathrm{UE}}$ serving UEs equipped with low-resolution ADCs. For a given synchronization time-slot, the BS simultaneously forms multiple analog synchronization beams, through which all potential UEs synchronize to the network.}
%\label{fig:figure5}
%\end{figure}
%By leveraging the design degrees of freedom in the spatial domain, we propose a multi-beam probing strategy of the synchronization signals under low-resolution ADCs.
In this section, we develop a new multi-beam probing based low-resolution timing synchronization strategy to effectively solve the multicast problem. Along with a common synchronization signal structure design, the proposed method exploits the spatial degrees of freedom arising from multiple RF chains to compensate for the quantization distortion without requiring explicit channel knowledge. Similar to Section III, we explain the proposed algorithm using the frequency-domain representations, though we first present the received signal model in the time-domain.

\subsection{Received signal model for multi-beam probing}
We assume that for a given synchronization time-slot, the BS deploys multiple subarrays to simultaneously form multiple analog synchronization beams, through which a total of $N_{\mathrm{UE}}$ UEs synchronize to the network. This is different from the conventional directional synchronization design in Section III-A, in which the BS probes one beam at a time. We set the corresponding baseband precoding matrix as identity matrix following the same methodology as in \cite{wonil,ttfv}. Further, for the given synchronization time-slot, we assume that the BS transmits common synchronization signals (or identical synchronization sequences) across the simultaneously probed beams. This is also different from conventional MIMO communications, in which distinct signals are spatially multiplexed to boost the capacity.

For a given synchronization time-slot, we denote the employed analog precoding matrix at the BS as
\begin{equation}\label{pcmtx}
\bm{P}=\left[
  \begin{array}{cccc}
    \bm{p}_{0} & \bm{0}_{N_{\mathrm{A}}\times1} & \cdots & \bm{0}_{N_{\mathrm{A}}\times1} \\
    \bm{0}_{N_{\mathrm{A}}\times1} & \bm{p}_{1} & \cdots & \bm{0}_{N_{\mathrm{A}}\times1} \\
    \vdots & \vdots & \ddots & \vdots \\
    \bm{0}_{N_{\mathrm{A}}\times1} & \bm{0}_{N_{\mathrm{A}}\times1} & \cdots & \bm{p}_{N_{\mathrm{RF}}-1} \\
  \end{array}
\right],
\end{equation}
where for $j=0,\cdots,N_{\mathrm{RF}}-1$, the beam probed from the $j$-th transmit RF chain $\bm{p}_{j}\in\mathbb{C}^{N_{\mathrm{A}}\times 1}$ satisfies the power constraint $\left[\bm{p}_{j}\bm{p}^{*}_{j}\right]_{a,a}=\frac{1}{N_{\mathrm{A}}}$ with $a=1,\cdots,N_{\mathrm{A}}$. Denote the set of the analog synchronization beams by $\Omega=\left\{\bm{p}_{0},\cdots,\bm{p}_{N_{\mathrm{RF}}-1}\right\}$. Similar to (\ref{zcmap16}) and (\ref{zcmap17}), we can then express the quantized time-domain received signal on the $b$-th receive antenna at UE $u\in\left\{1,\cdots,N_{\mathrm{UE}}\right\}$ as
\begin{eqnarray}
q^{\Omega}_{u,b}[\mathrm{t}+n]&=&\mathcal{Q}\left(y^{\Omega}_{u,b}[\mathrm{t}+n]\right)\\
&=&\mathcal{Q}\left(\left[\sum_{\ell=0}^{L_u-1}\sqrt{\frac{1}{N_{\mathrm{RF}}}}\bm{H}_u[\ell]\bm{P}\left(d[((k-\ell))_{N}]\bm{1}_{N_{\mathrm{RF}}\times1}\right)\right]_{b}+w_u[n]\right)\label{vecneed}\\
&=&\mathcal{Q}\Bigg(\sum_{\ell=0}^{L_u-1}\sum_{j=0}^{N_{\mathrm{RF}}-1}\sqrt{\frac{1}{N_{\mathrm{RF}}}}\left(\left[\bm{H}_u[\ell]\right]_{b,jN_{\mathrm{A}}+1:(j+1)N_{\mathrm{A}}}\bm{p}_{j}\right)d[((n-\ell))_{N}]+w_u[n]\Bigg),\nonumber\\ \label{39}
\end{eqnarray}
where the transmit power is scaled by the number of streams, i.e., $N_{\mathrm{RF}}$, to maintain the total power constraint. Note that $\bm{1}_{N_{\mathrm{RF}}\times 1}$ in (37) indicates the common synchronization signal structure.

To express (\ref{39}) in vector form, we first denote $h^{\Omega}_{u,b}[\ell]=\sum_{j=0}^{N_{\mathrm{RF}}-1}\left[\bm{H}_u[\ell]\right]_{b,jN_{\mathrm{A}}+1:(j+1)N_{\mathrm{A}}}\bm{p}_{j}$ as the time-domain composite effective transmit beam-space channel and $\bm{q}^{\Omega}_{u,b}=\big[q^{\Omega}_{u,b}[\mathrm{t}],\cdots,q^{\Omega}_{u,b}[\mathrm{t}+N-1]\big]^{\mathrm{T}}$. By applying Bussgang's theorem, we then have
\begin{eqnarray}\label{similar}
\bm{q}^{\Omega}_{u,b}=\bm{E}^{\Omega}_{u,b}\Bigg(\underbrace{\sum_{\ell=0}^{L_u-1}\sqrt{\frac{1}{N_{\mathrm{RF}}}}h^{\Omega}_{u,b}[\ell]\bm{d}_{\ell}}_{\bm{v}^{\Omega}_{u,b}}+\bm{w}_u\Bigg)+\check{\bm{w}}^{\Omega}_{u,b},
\end{eqnarray}
where the quantization distortion matrix $\bm{E}^{\Omega}_{u,b}=\mathrm{diag}\left(\left[\eta^{\Omega}_{u,b}[0],\cdots,\eta^{\Omega}_{u,b}[N-1]\right]^{\mathrm{T}}\right)$, and similar to (\ref{defnmse}), the corresponding quantization distortion factor is
\begin{equation}\label{accpt}
\eta^{\Omega}_{u,b}[n] = \frac{\mathbb{E}\left[(q^{\Omega}_{u,b}[\mathrm{t}+n])^{*}y^{\Omega}_{u,b}[\mathrm{t}+n]\right]}{\mathbb{E}\left[\left|y^{\Omega}_{u,b}[\mathrm{t}+n]\right|^2\right]}.
\end{equation}
Denoting the covariance matrix of the noiseless unquantized received signal $\bm{v}^{\Omega}_{u,b}$ in (\ref{similar}) by $\bm{R}_{\bm{v}^{\Omega}_{u,b}}$, we can further express $\bm{E}^{\Omega}_{u,b}$ as
\begin{equation}\label{new00}
\bm{E}^{\Omega}_{u,b}=(1-\xi_u)\mathrm{diag}\left(\bm{R}_{\bm{v}^{\Omega}_{u,b}}+\sigma^2\bm{I}_{N}\right)^{-\frac{1}{2}}.
\end{equation}
The covariance matrix of the quantization noise vector $\check{\bm{w}}^{\Omega}_{u,b}$ with respect to the $b$-th receive antenna at UE $u$ now becomes
\begin{eqnarray}\label{newcov}
\bm{R}_{\check{\bm{w}}^{\Omega}_{u,b}}=\bm{E}^{\Omega}_{u,b}\left(\bm{I}_{N}-\bm{E}^{\Omega}_{u,b}\right)\underbrace{\mathrm{diag}\left(\bm{R}_{\bm{v}^{\Omega}_{u,b}}+\sigma^2\bm{I}_{N}\right)}_{\bm{V}^{\Omega}_{u,b}},
\end{eqnarray}
where $\bm{V}^{\Omega}_{u,b}$ represents the corresponding unquantized received signal power matrix.

\subsection{Optimization problem formulation for multi-beam probing}
Prior to formulating the low-resolution timing synchronization problem for the proposed multi-beam probing, we need to first derive the correlation between the frequency-domain quantized received synchronization signal $\textbf{\textsf{q}}^{\Omega}_{u,b}$ and the known unquantized reference synchronization sequence $\textbf{\textsf{d}}$. Denoting
\begin{equation}\label{cetbch}
\textsf{h}^{\Omega}_{u,b}[k] = \sum_{j=0}^{N_{\mathrm{RF}}-1}\left[\bm{A}^{\mathrm{RX}}_{u}\textbf{\textsf{G}}_u[k]\left(\bm{A}^{\mathrm{TX}}_{u}\right)^{*}\right]_{b,jN_{\mathrm{A}}+1:(j+1)N_{\mathrm{A}}}\bm{p}_{j},
\end{equation}
as the frequency-domain composite effective transmit beam-space channel relative to its time-domain counterpart $h^{\Omega}_{u,b}[\ell]$, we can compute the zero-lag frequency-domain correlation for UE $u$ as
\begin{align}
\Lambda^{\Omega}_{u,b}[0] &= \sum_{k=0}^{N-1}\textsf{q}^{\Omega}_{u,b}[k]\textsf{d}^{*}[k]\\
&=\sum_{k=0}^{N-1}\eta^{\Omega}_{u,b}[k]\sqrt{\frac{1}{N_{\mathrm{RF}}}}\textsf{h}^{\Omega}_{u,b}[k]\textsf{d}[k]\textsf{d}^{*}[k]+\sum_{k=0}^{N-1}\eta^{\Omega}_{u,b}[k]\textsf{w}_u[k]\textsf{d}^{*}[k] + \sum_{k=0}^{N-1}\check{\textsf{w}}^{\Omega}_{u,b}[k]\textsf{d}^{*}[k].
\end{align}
Denoting by $\hat{b}=\underset{b=1,\cdots,M_{\mathrm{tot}}}{\mathrm{argmax}}\left|\Lambda^{\Omega}_{u,b}[0]\right|^2$ and applying the same flat synchronization channels assumption as in (\ref{snqr0}), we can rewrite the frequency-domain composite effective transmit beam-space channel in (\ref{cetbch}) as
\begin{equation}\label{ttt}
\textsf{h}^{\Omega}_{u} = \sum_{j=0}^{N_{\mathrm{RF}}-1}\left[\bm{a}^{*}_{\mathrm{tx}}(\theta_u,\phi_u)\right]_{jN_{\mathrm{A}}+1:(j+1)N_{\mathrm{A}}}\bm{p}_{j}.
\end{equation}
Further, denoting the frequency-domain counterpart of the unquantized received signal power matrix $\bm{V}^{\Omega}_{u,\hat{b}}$ in (\ref{newcov}) by $\textbf{\textsf{V}}^{\Omega}_{u,\hat{b}}$, we have
\begin{eqnarray}\label{new01}
\textbf{\textsf{V}}^{\Omega}_{u,\hat{b}}&=&\frac{\textsf{g}^{2}_{u}\left|\left[\bm{a}_{\mathrm{rx}}(\psi_u)\right]_{\hat{b}}\right|^2}{N_{\mathrm{RF}}}\left|\textsf{h}^{\Omega}_{u}\right|^2\mathrm{diag}\left(\left[\textsf{d}[0]\textsf{d}^{*}[0],\cdots,\textsf{d}[N-1]\textsf{d}^{*}[N-1]\right]^{\mathrm{T}}\right)+\sigma^2\bm{I}_{N}.
\end{eqnarray}
By exploiting the inherent correlation properties of the common synchronization signal design and applying the common quantization distortion factor $\underline{\eta}^{\Omega}_{u,\hat{b}}=\eta^{\Omega}_{u,\hat{b}}[0]=\cdots=\eta^{\Omega}_{u,\hat{b}}[N-1]$, we can obtain the quantization noise power by using (\ref{new00}), (\ref{newcov}) and (\ref{new01}), which results in the received synchronization SQNR at zero-lag correlation for UE $u$ as
\begin{equation}\label{snqr1}
\gamma^{\Omega}_{u,\hat{b}}=\frac{\underline{\eta}^{\Omega}_{u,\hat{b}}\frac{\textsf{g}^{2}_{u}\left|\left[\bm{a}_{\mathrm{rx}}(\psi_u)\right]_{\hat{b}}\right|^2}{N_{\mathrm{RF}}}\left|\textsf{h}^{\Omega}_{u}\right|^2}{\underline{\eta}^{\Omega}_{u,\hat{b}}\sigma^2+\left(1-\underline{\eta}^{\Omega}_{u,\hat{b}}\right)\left(\frac{\textsf{g}^{2}_{u}\left|\left[\bm{a}_{\mathrm{rx}}(\psi_u)\right]_{\hat{b}}\right|^2}{N_{\mathrm{RF}}}\left|\textsf{h}^{\Omega}_{u}\right|^2+\sigma^2\right)}.
\end{equation}
For the given synchronization time-slot, we formulate the corresponding max-min optimization problem for the proposed multi-beam probing based multicast as
\begin{eqnarray}\label{oringmaxminopnewrev}
&\mathcal{P}4:\hspace{2mm}\underset{\Omega}{\mathrm{max}}\hspace{2mm}\underset{\forall u}{\mathrm{min}}\left\{\gamma^{\Omega}_{u,\hat{b}}\right\}&\\
&\textrm{s.t.}\hspace{2mm}\Omega\in\left(\mathcal{F}\right)^{N_{\mathrm{RF}}}.&\nonumber
\end{eqnarray}
Denote by $\lambda'_{u}=N_{\mathrm{RF}}\sigma^2/\textsf{g}^{2}_{u}\left|\left[\bm{a}_{\mathrm{rx}}(\psi_u)\right]_{\hat{b}}\right|^2$ for $u=1,\cdots,N_{\mathrm{UE}}$ and $\lambda'_{\mathrm{max}}=\mathrm{max}\left\{\lambda'_1,\cdots,\lambda'_{N_{\mathrm{UE}}}\right\}$. Similar to (\ref{maxminone}), we obtain a lower bound of $\gamma^{\Omega}_{u,\hat{b}}$ as
\begin{equation}\label{rrrttt}
\acute{\gamma}^{\Omega}_{u}=\frac{\left|\textsf{h}^{\Omega}_{u}\right|^2}{\lambda'_{\mathrm{max}}+\left[\frac{\left[\sigma^2\left(\left|\textsf{h}^{\Omega}_{u}\right|^{2}/\lambda'_{\mathrm{max}}+1\right)\right]^{1/2}}{1-\xi_{\mathrm{max}}}-1\right]\left(\left|\textsf{h}^{\Omega}_{u}\right|^2+\lambda'_{\mathrm{max}}\right)},
\end{equation}
such that
\begin{equation}\label{lessbound2}
\acute{\gamma}^{\Omega}_{u}\leq\gamma^{\Omega}_{u,\hat{b}},\hspace{2mm}u\in\left\{1,\cdots,N_{\mathrm{UE}}\right\}.
\end{equation}
Based on (\ref{lessbound2}), we reformulate (\ref{oringmaxminopnewrev}) as
\begin{eqnarray}\label{realmaxminopnewrev}
&\mathcal{P}5:\hspace{2mm}\underset{\Omega}{\mathrm{max}}\hspace{2mm}\underset{\forall u}{\mathrm{min}}\left\{\acute{\gamma}^{\Omega}_{u}\right\}&\\
&\textrm{s.t.}\hspace{2mm}\Omega\in\left(\mathcal{F}\right)^{N_{\mathrm{RF}}}.&\nonumber
\end{eqnarray}
\subsection{Proposed multi-beam probing based low-resolution synchronization design}
\begin{figure}
\centering
\includegraphics[width=5.8in]{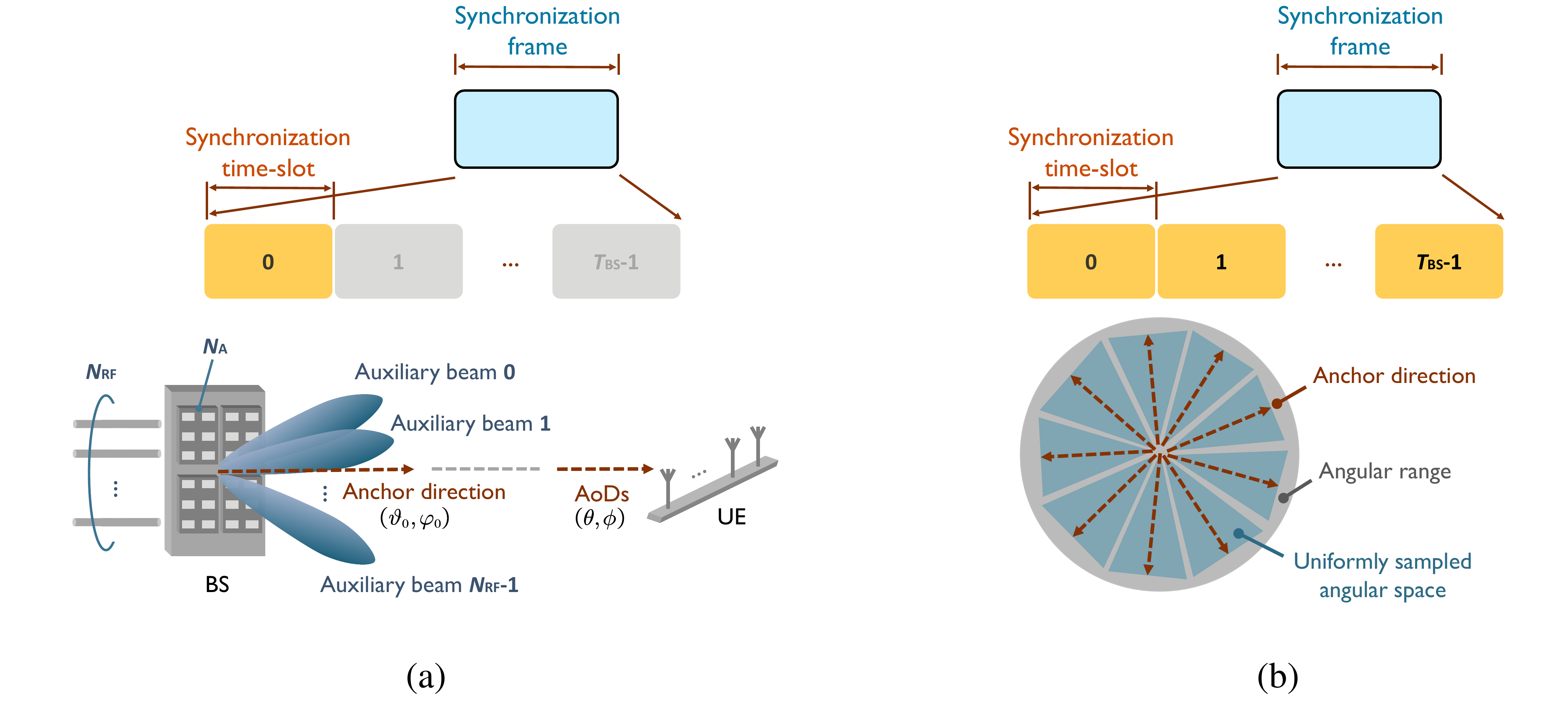}
\caption{(a) Conceptual examples of anchor direction and auxiliary beams for synchronization time-slot $0$. (b) Across $T_{\mathrm{BS}}$ synchronization time-slots, all $T_{\mathrm{BS}}$ anchor directions uniformly sample the given angular space.}
\label{fig:figure6}
\end{figure}

Similar to $\mathcal{P}3$ in (\ref{realmaxminoprev}), solving $\mathcal{P}5$ in (\ref{realmaxminopnewrev}) also requires the BS to have explicit knowledge of the channel directional information of all UEs, which is unavailable during the initial frame timing synchronization phase. In our proposed design, the UEs with significantly different AoDs are not supposed to synchronize to the network simultaneously, but instead via different synchronization time-slots. That is, for a given synchronization time-slot, we are only interested in a certain group of UEs that have similar AoDs. We therefore define an anchor direction to characterize those similar AoDs as much as possible for the synchronization time-slot of interest. As one synchronization frame contains $T_{\mathrm{BS}}$ synchronization time-slots, we define a set of $T_{\mathrm{BS}}$ anchor angular directions to represent the potential channel directions. We provide an example of the anchor direction for a given synchronization time-slot in Fig.~\ref{fig:figure6}(a). The anchor directions and $T_{\mathrm{BS}}$ synchronization time-slots have one-to-one mapping, and all $T_{\mathrm{BS}}$ anchor directions uniformly sample the angular range of interest, which is shown in Fig.~\ref{fig:figure6}(b). If the number of synchronization time-slots $T_{\mathrm{BS}}\rightarrow\infty$, the anchor directions fully sample the given angular space such that they characterize all possible channel directions. Note that other choices of the anchor directions are possible (e.g., non-uniformly sample the given angular space), depending on practical system requirements.

Based on the definition of anchor direction, we now reformulate the optimization problem in (\ref{realmaxminopnewrev}). We first define $\vartheta^{\star}$ and $\varphi^{\star}$ as the azimuth and elevation anchor directions for the synchronization time-slot of interest and use them to represent the potential channel's azimuth and elevation AoDs. Note that this representation becomes more accurate as $T_{\mathrm{BS}}$ increases. Similar to (\ref{ttt}), we first define
\begin{eqnarray}\label{qqq}
\textsf{h}^{\Omega^{\star}} = \sum_{j=0}^{N_{\mathrm{RF}}-1}\left[\bm{a}^{*}_{\mathrm{tx}}\left(\vartheta^{\star}+\varphi^{\star}\right)\right]_{jN_{\mathrm{A}}+1:(j+1)N_{\mathrm{A}}}\bm{p}_{j},
\end{eqnarray}
where $\Omega^{\star}$ denotes the set of candidate beams with $\vartheta^{\star}$ and $\varphi^{\star}$ as the corresponding azimuth and elevation anchor directions. Similar to (\ref{rrrttt}), we can then obtain
\begin{equation}
\acute{\gamma}^{\Omega^{\star}}=\frac{\left|\textsf{h}^{\Omega^{\star}}\right|^2}{\lambda'_{\mathrm{max}}+\left[\frac{\left[\sigma^2\left(\left|\textsf{h}^{\Omega^{\star}}\right|^{2}/\lambda'_{\mathrm{max}}+1\right)\right]^{1/2}}{1-\xi_{\mathrm{max}}}-1\right]\left(\left|\textsf{h}^{\Omega^{\star}}\right|^2+\lambda'_{\mathrm{max}}\right)}.
\end{equation}
\begin{figure}
\begin{center}
\subfigure[]{%
\includegraphics[width=2.4in]{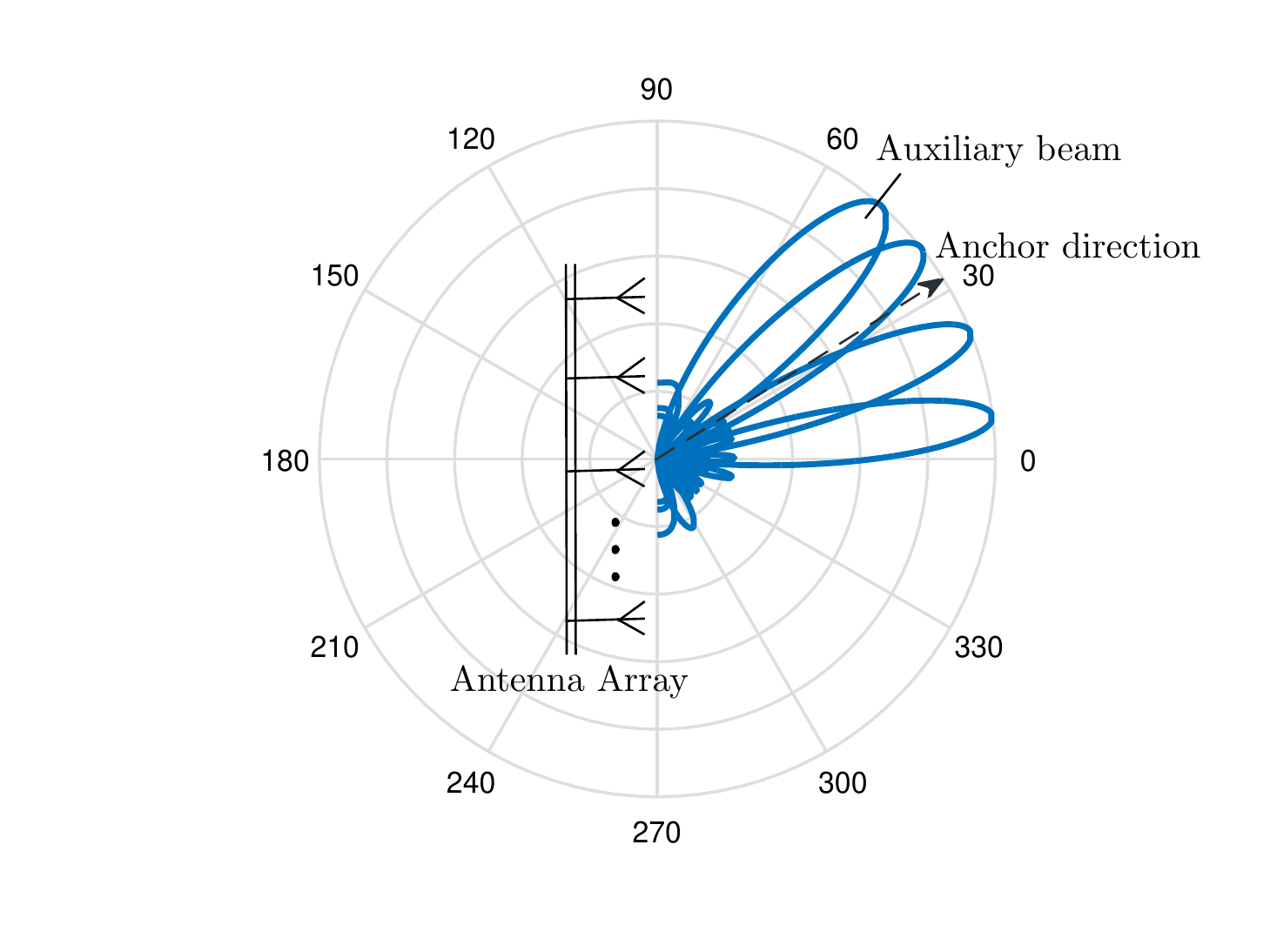}
\label{fig:subfigure7a}}
\subfigure[]{%
\includegraphics[width=2.4in]{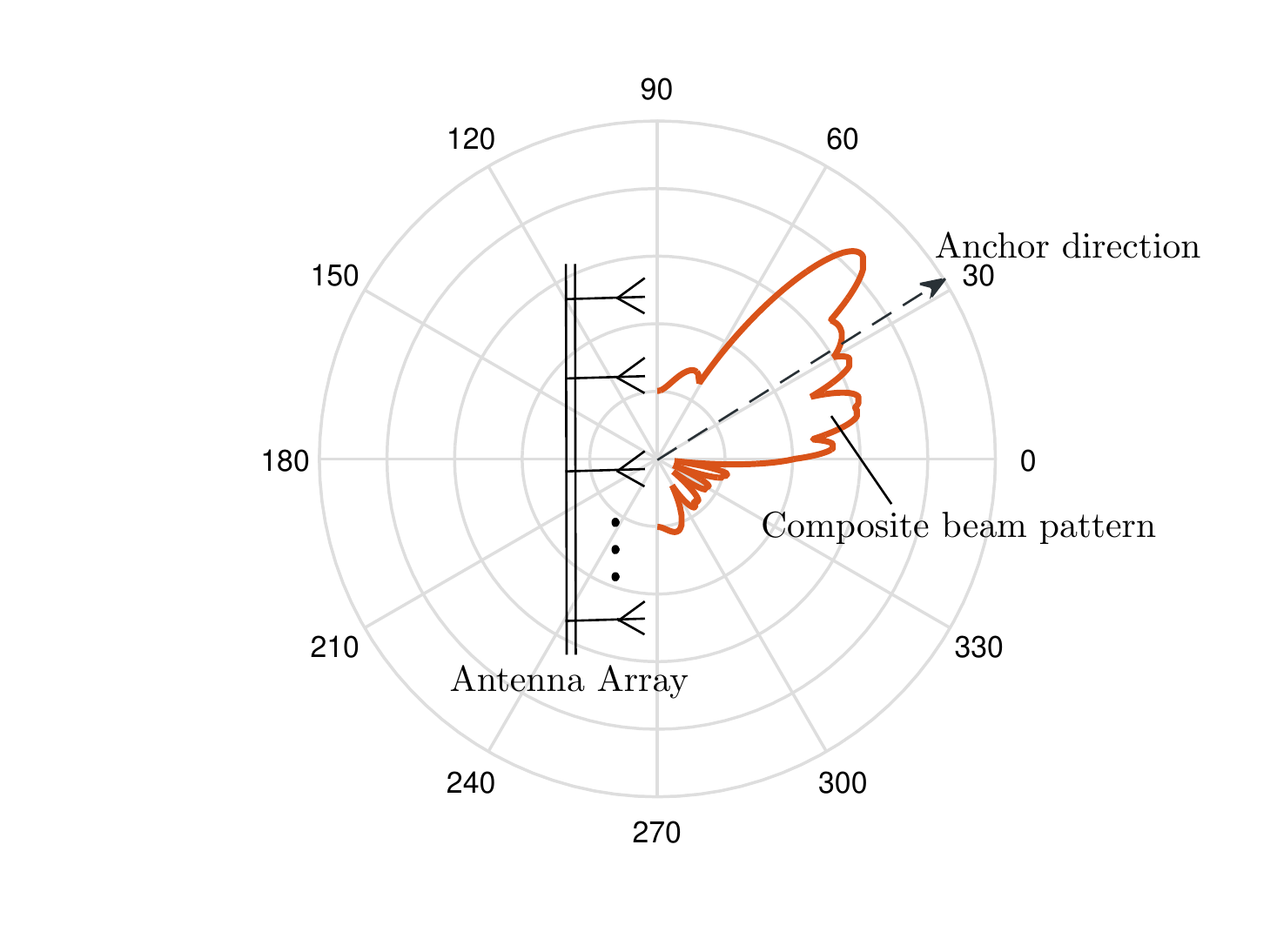}
\label{fig:subfigure7b}}
%\subfigure[]{%
%\includegraphics[width=2.6in]{BeamPatternIndependent1.pdf}
%\label{fig:subfigure7c}}
%\subfigure[]{%
%\includegraphics[width=2.6in]{BeamPatternComposite1.pdf}
%\label{fig:subfigure7d}}
\subfigure[]{%
\includegraphics[width=2.4in]{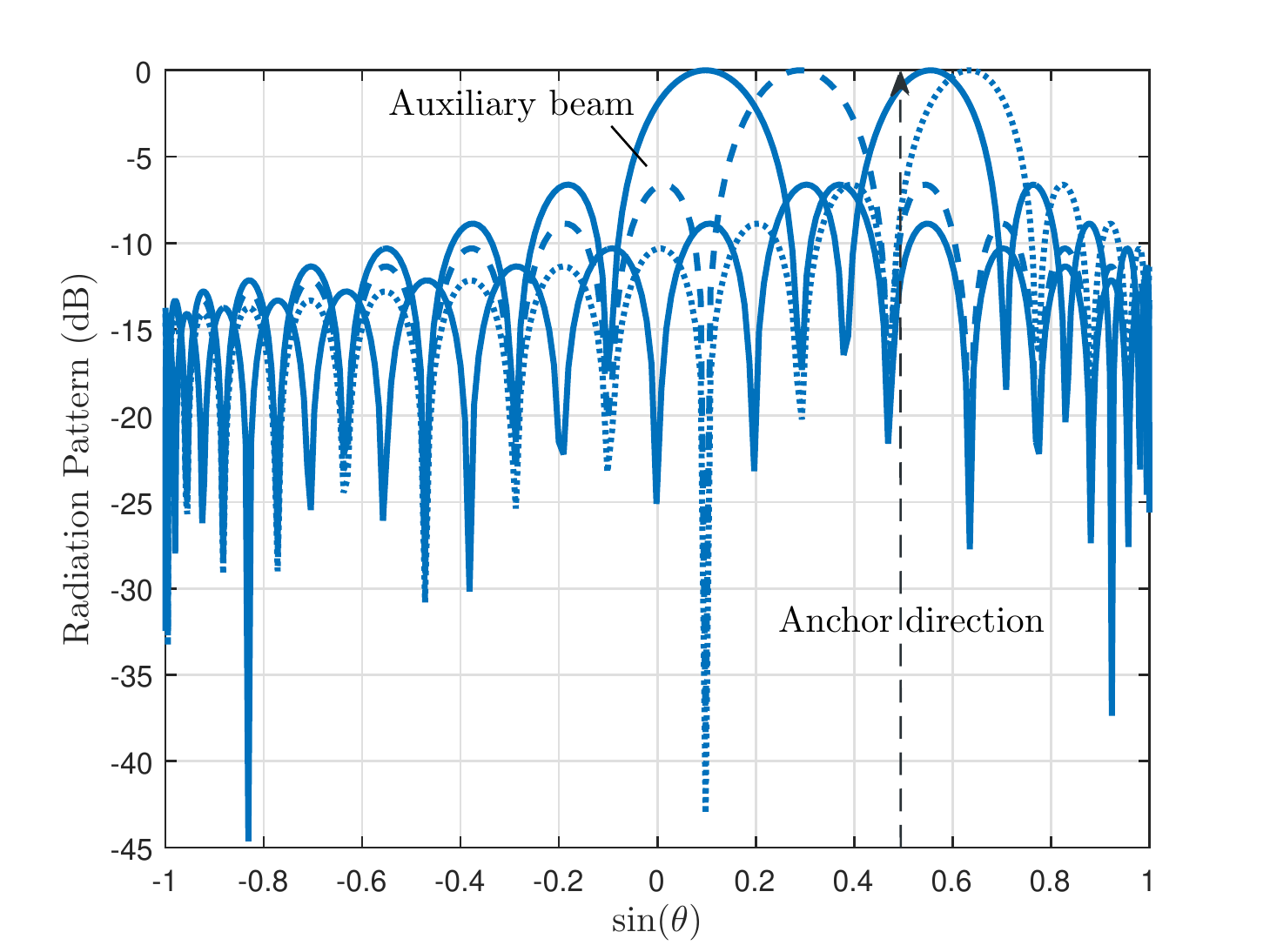}
\label{fig:subfigure7e}}
\subfigure[]{%
\includegraphics[width=2.4in]{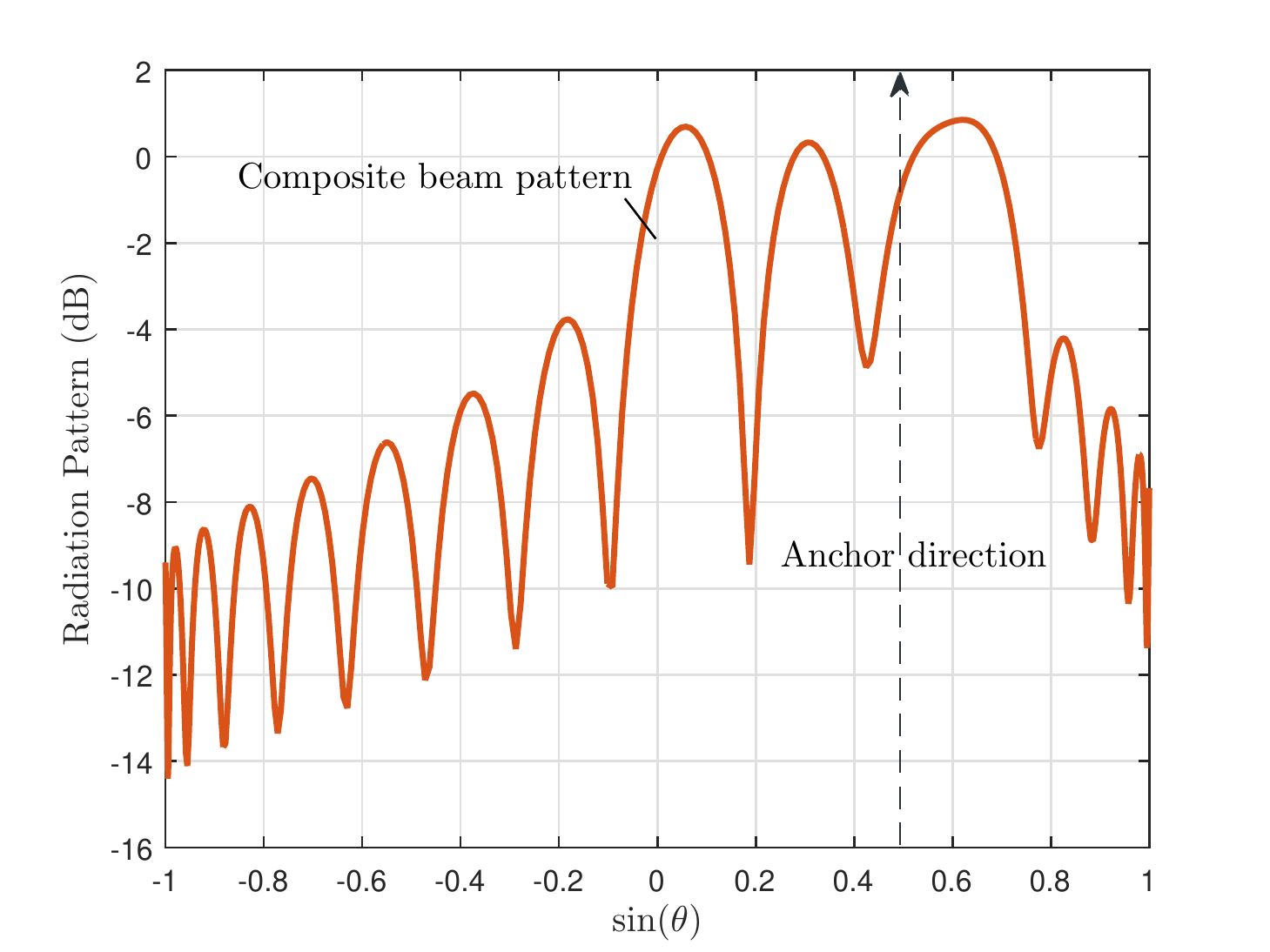}
\label{fig:subfigure7f}}
\caption{An example of auxiliary beams is provided in (a); the corresponding composite beam is presented in (b). Radiation patterns of the auxiliary beams and the composite beam in (a) and (b) are plotted in (c) and (d). A ULA is assumed with $N_{\mathrm{tot}}=8$ and $N_{\mathrm{RF}}=4$. A DFT beam codebook with oversampling factor of $2$ is employed.}
\label{fig:figure7abcdef}
\end{center}
\end{figure}
According to (\ref{realmaxminopnewrev}), we formulate the optimization problem as
\begin{eqnarray}\label{profh}
&\mathcal{P}6:\hspace{2mm}\underset{\Omega^{\star}}{\mathrm{max}}\left\{\acute{\gamma}^{\Omega^{\star}}\right\}&\\
&\textrm{s.t.}\hspace{2mm}\Omega^{\star}\in\left(\mathcal{F}\right)^{N_{\mathrm{RF}}},&\nonumber
\end{eqnarray}
which transforms the complex max-min optimization problem into a maximization problem. To solve (\ref{profh}), the BS can execute the exhaustive search over all possible combinations among the candidate beam codewords in $\mathcal{F}$, resulting in
\begin{equation}\label{optsim}
\Omega^{\star}_{\mathrm{opt}} = \underset{\Omega^{\star}\in\left(\mathcal{F}\right)^{N_{\mathrm{RF}}}}{\mathrm{argmax}} \left\{\acute{\gamma}^{\Omega^{\star}}\right\},
\end{equation}
for the given synchronization time-slot. We refer to the simultaneously probed beams in $\Omega^{\star}_{\mathrm{opt}}$ as auxiliary beams for the synchronization time-slot of interest as depicted in Fig.~\ref{fig:figure6}(a).
%We now explain (\ref{optsim}) using the conceptual example provided in Fig.~\ref{fig:figure6}. Building on the definition of anchor direction, we denote the simultaneously probed beams in $\Omega^{\star}_{\mathrm{opt}}$ as auxiliary beams for the synchronization time-slot of interest. Consider synchronization time-slot $0$ with the corresponding azimuth and elevation anchor directions $\vartheta_0$ and $\varphi_0$ in Fig.~\ref{fig:figure6}(a). The main target of forming the auxiliary beams for synchronization time-slot $0$ is to ensure that if a given UE's azimuth and elevation AoDs $\theta$ and $\phi$ are best characterized by $\vartheta_0$ and $\varphi_0$, i.e., $\left|\vartheta_0-\theta\right|=\min\left\{\left|\vartheta_t-\theta\right|,t=0,\cdots,T_{\mathrm{BS}}-1\right\}$ and $\left|\varphi_0-\phi\right|=\min\left\{\left|\varphi_t-\phi\right|,t=0,\cdots,T_{\mathrm{BS}}-1\right\}$, the UE would experience the highest received synchronization SQNR at zero-lag correlation when accessing to synchronization time-slot $0$ among all synchronization time-slots (depicted by a conceptual plot in Fig.~\ref{fig:figure6}(c)).

From (\ref{cetbch}), (\ref{ttt}) and (\ref{qqq}), for the proposed multi-beam probing strategy with common synchronization signal structure, the simultaneously probed auxiliary beams actually form an effective composite beam. In Fig.~\ref{fig:figure7abcdef}, we provide examples of the auxiliary beams and the corresponding effective composite beam pattern. The auxiliary beams are selected from a DFT beam codebook with oversampling factor of $2$. As evident from Figs.~\ref{fig:subfigure7a} and \ref{fig:subfigure7b}, the effective composite beam may not yield the largest beamforming gain towards the corresponding anchor direction. In Figs.~\ref{fig:subfigure7e} and \ref{fig:subfigure7f}, we plot the radiation patterns corresponding to the auxiliary beams and the composite beam in Figs.~\ref{fig:subfigure7a} and \ref{fig:subfigure7b}. In summary, by optimizing the effective composite beam pattern via (\ref{optsim}), we optimize the distribution/resolution of the input samples to the quantizer such that a better tradeoff between the beamforming gain and the resulted quantization distortion can be achieved.
%\subsection{Practical implementation issues}
%As can be seen from Fig.~7, the effective composite beam patterns have relatively large spatial side-lobes, which may result in high inter/intra-cell interference. For synchronization channels beamforming, the neighboring cells usually employ different synchronization sequences with different root indices, which would eliminate the inter-cell interference because of the orthogonality in the code domain. Further, for a given cell, the proposed multi-beam probing is conducted in a TDM round-robin fashion across all synchronization time-slots, which avoids the intra-cell interference. Hence, the proposed directional synchronization design does not introduce any new interference scenario.
%
%The proposed method focuses on the BS side processing by optimizing the corresponding beamforming strategy without bringing extra implementation complexity to the low-resolution receivers. Upon receiving the synchronization signals, the UE with low-resolution ADCs simply correlates the received signal samples with the local reference synchronization sequence and detects the peak power.
\subsection{Computational complexity for multi-beam probing}
From the perspective of the BS, the proposed multi-beam probing involves beam search over a given beam codebook $\mathcal{F}$ across a total of $N_{\mathrm{RF}}$ transmit RF chains. Denote the number of beam codewords in $\mathcal{F}$ by $N_{\mathrm{beam}}$. For a given synchronization time-slot, selecting appropriate $N_{\mathrm{RF}}$ synchronization beams according to (\ref{optsim}) requires $\left(N_{\mathrm{beam}}\right)^{N_{\mathrm{RF}}}$ iterations for all possible choices. For a total of $T_{\mathrm{BS}}$ synchronization time-slots, this number becomes $T_{\mathrm{BS}}\left(N_{\mathrm{beam}}\right)^{N_{\mathrm{RF}}}$. If the whole optimization process is triggered $N_{\mathrm{T}}$ times, the overall computational complexity for our proposed design is $N_{\mathrm{T}}T_{\mathrm{BS}}\left(N_{\mathrm{beam}}\right)^{N_{\mathrm{RF}}}$, which is larger than $N_{\mathrm{T}}T_{\mathrm{BS}}N_{\mathrm{beam}}$ for conventional single-stream beamforming based approaches. In this paper, we assume that the BS conducts the beam optimization in a semi-static manner, and the resulted access delay is negligible.

From the perspective of the UE, the corresponding computational complexity mainly comes from correlating the received signal samples with the locally stored synchronization sequence according to (\ref{crosscorr}). Implementing (\ref{crosscorr}) requires $(N+1)$ complex multiplication and $(N-1)$ complex addition operations. For all values of $\nu$ in (\ref{crosscorr}), the total number of operations is $N(N+1)(T_{\mathrm{UE}}-1)$ complex multiplication and $N(N-1)(T_{\mathrm{UE}}-1)$ complex addition operations. Across all $M_{\mathrm{tot}}$ receive antennas, the total number of complex multiplication and addition operations becomes $M_{\mathrm{tot}}N(N+1)(T_{\mathrm{UE}}-1)$ and $M_{\mathrm{tot}}N(N-1)(T_{\mathrm{UE}}-1)$. Note that the number of complex multiplication and addition operations is the same for both the proposed multi-beam probing and conventional single-stream beamforming based designs. That is, the proposed approach does not introduce additional implementation complexity to the UE.
\section{Numerical Results}
\begin{figure}
\begin{center}
\subfigure[]{%
\includegraphics[width=2.6in]{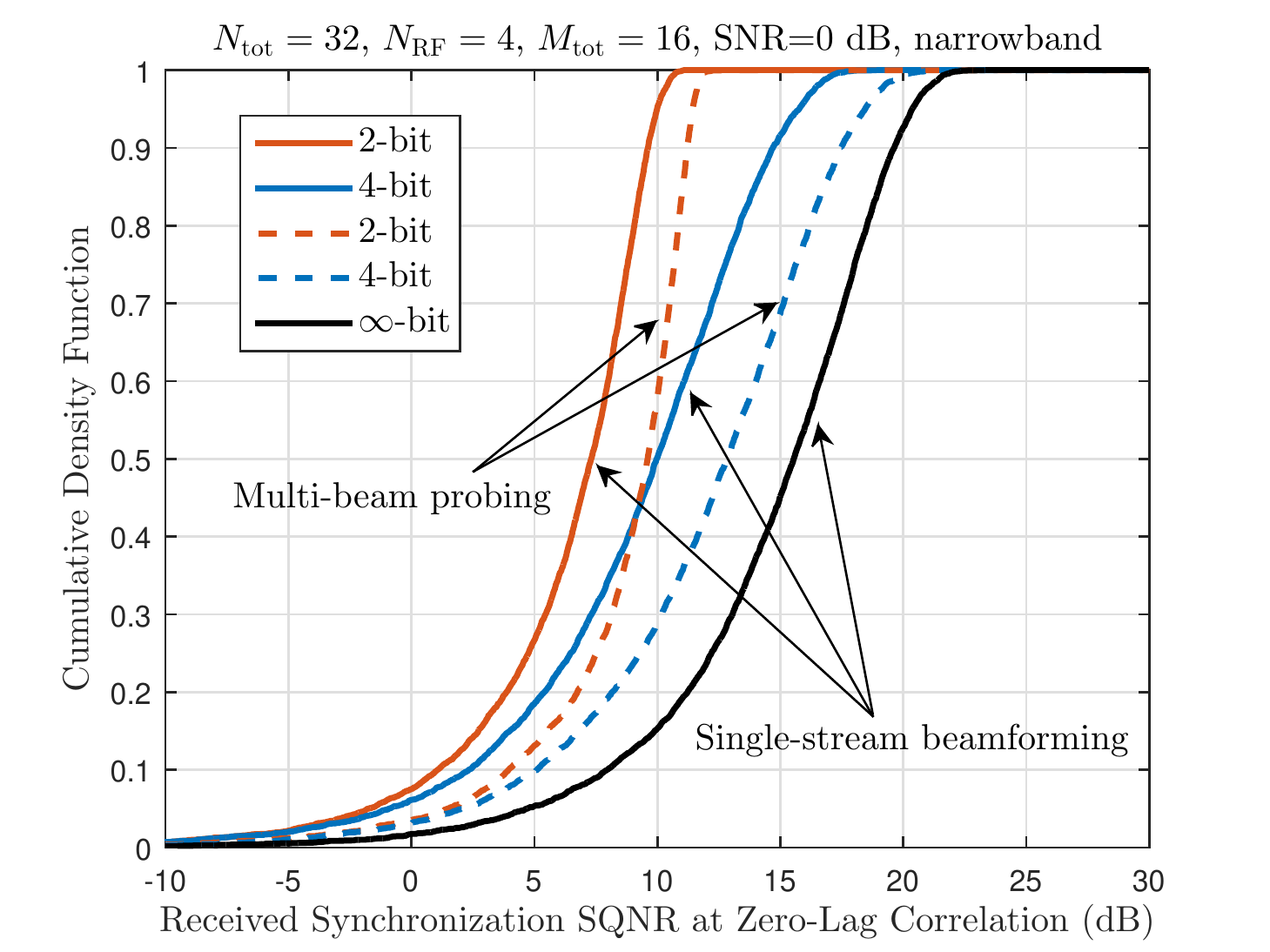}
\label{fig:subfigure8a}}
%\hspace{-3.5mm}
\subfigure[]{%
\includegraphics[width=2.6in]{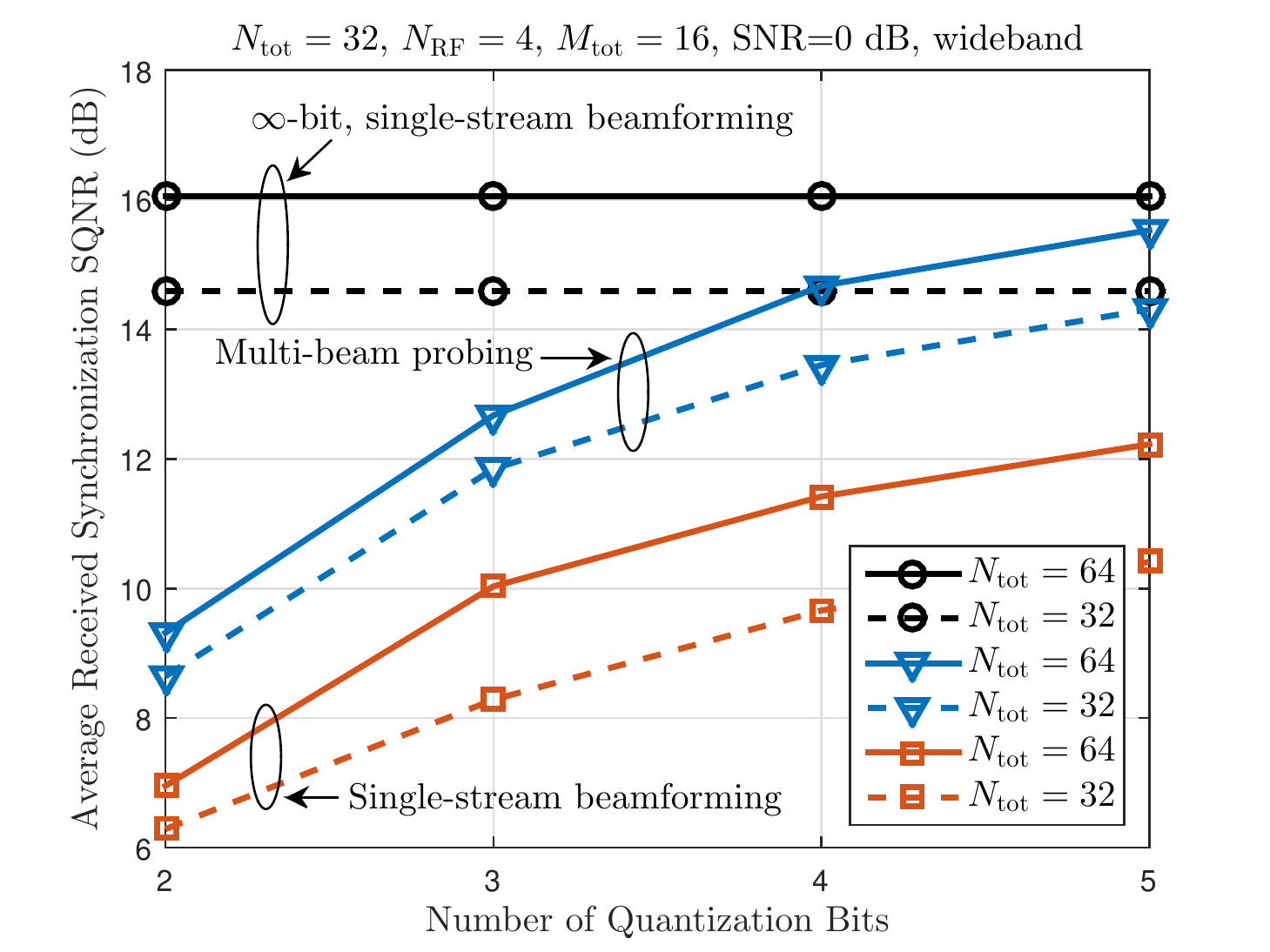}
\label{fig:subfigure8b}}
\caption{(a) CDFs of the received synchronization SQNRs in single-path frequency-flat channels; $N_{\mathrm{tot}}=32$ with $N_{\mathrm{RF}}=4$. (b) Average received synchronization SQNRs versus the number of quantization bits in multi-path frequency-selective channels as in \cite{kssr}; $N_{\mathrm{tot}}=32$ with $N_{\mathrm{RF}}=4$.}
\label{fig:figure8ab}
\end{center}
\end{figure}

\begin{figure}
\begin{center}
\subfigure[]{%
\includegraphics[width=2.6in]{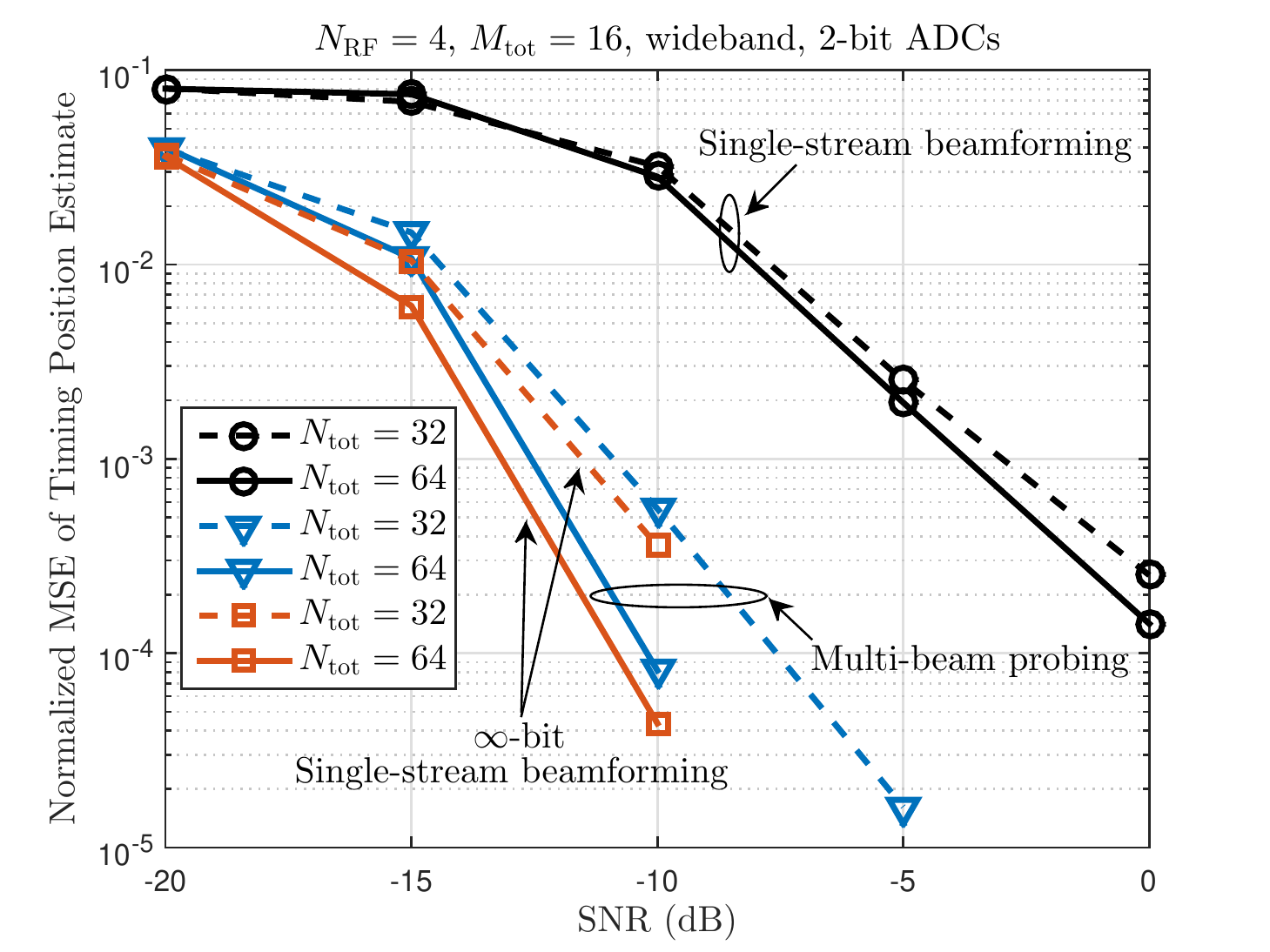}
\label{fig:subfigure9a}}
%\hspace{-3.5mm}
\subfigure[]{%
\includegraphics[width=2.6in]{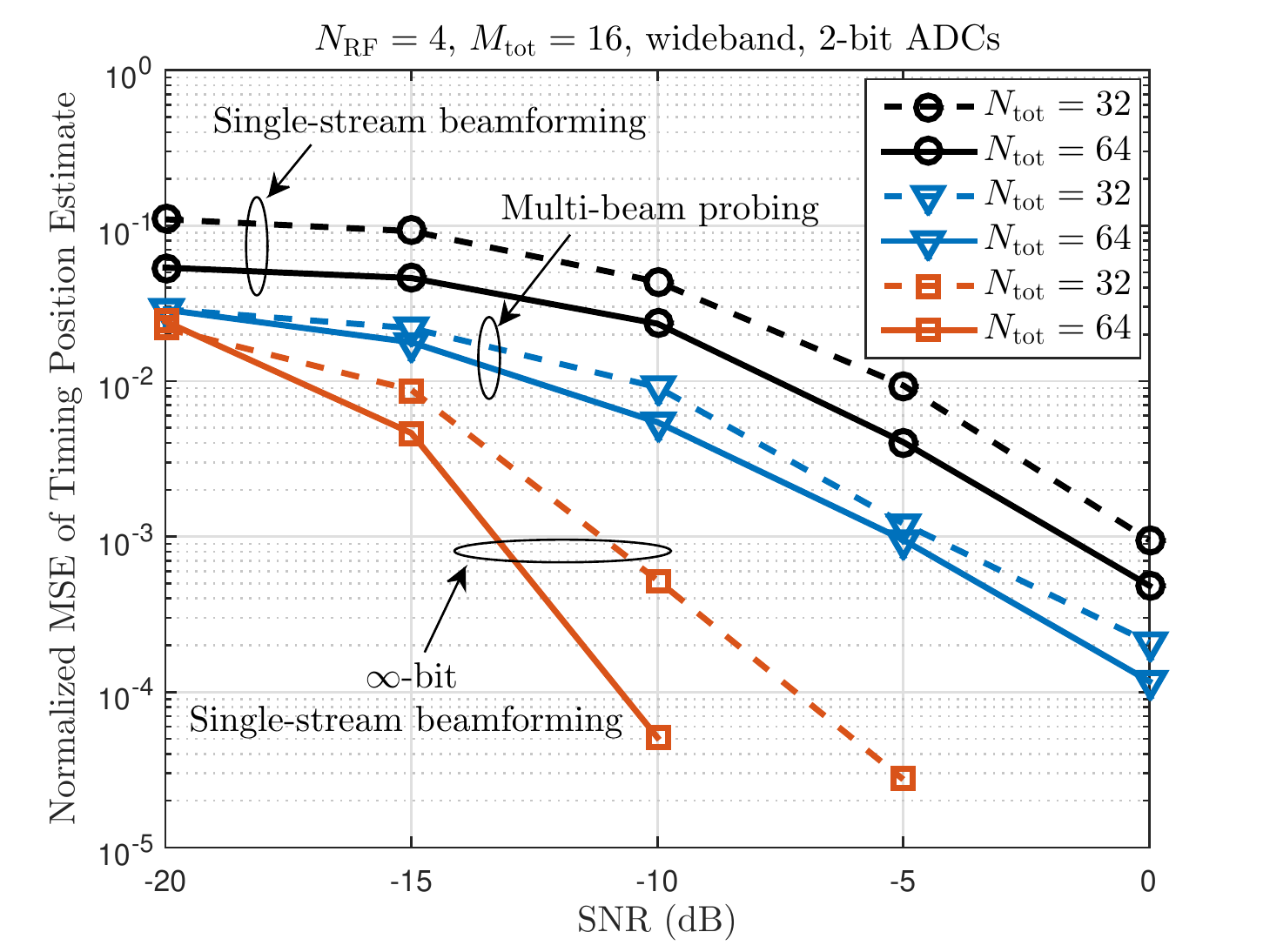}
\label{fig:subfigure9b}}
\caption{(a) Normalized MSE of timing position estimate versus SNR. Multi-path frequency-selective channels are employed assuming a single UE equipped with a $2$-bit ADCs receiver; $N_{\mathrm{tot}}=\{32,64\}$ with $N_{\mathrm{RF}}=4$. (b) Normalized MSE of timing position estimate versus SNR. Multi-path frequency-selective channels are employed assuming multiple UEs equipped with $2$-bit ADCs receivers; $N_{\mathrm{tot}}=\{32,64\}$ with $N_{\mathrm{RF}}=4$.}
\label{fig:figure9ab}
\end{center}
\end{figure}

\begin{figure}
\begin{center}
\subfigure[]{%
\includegraphics[width=2.6in]{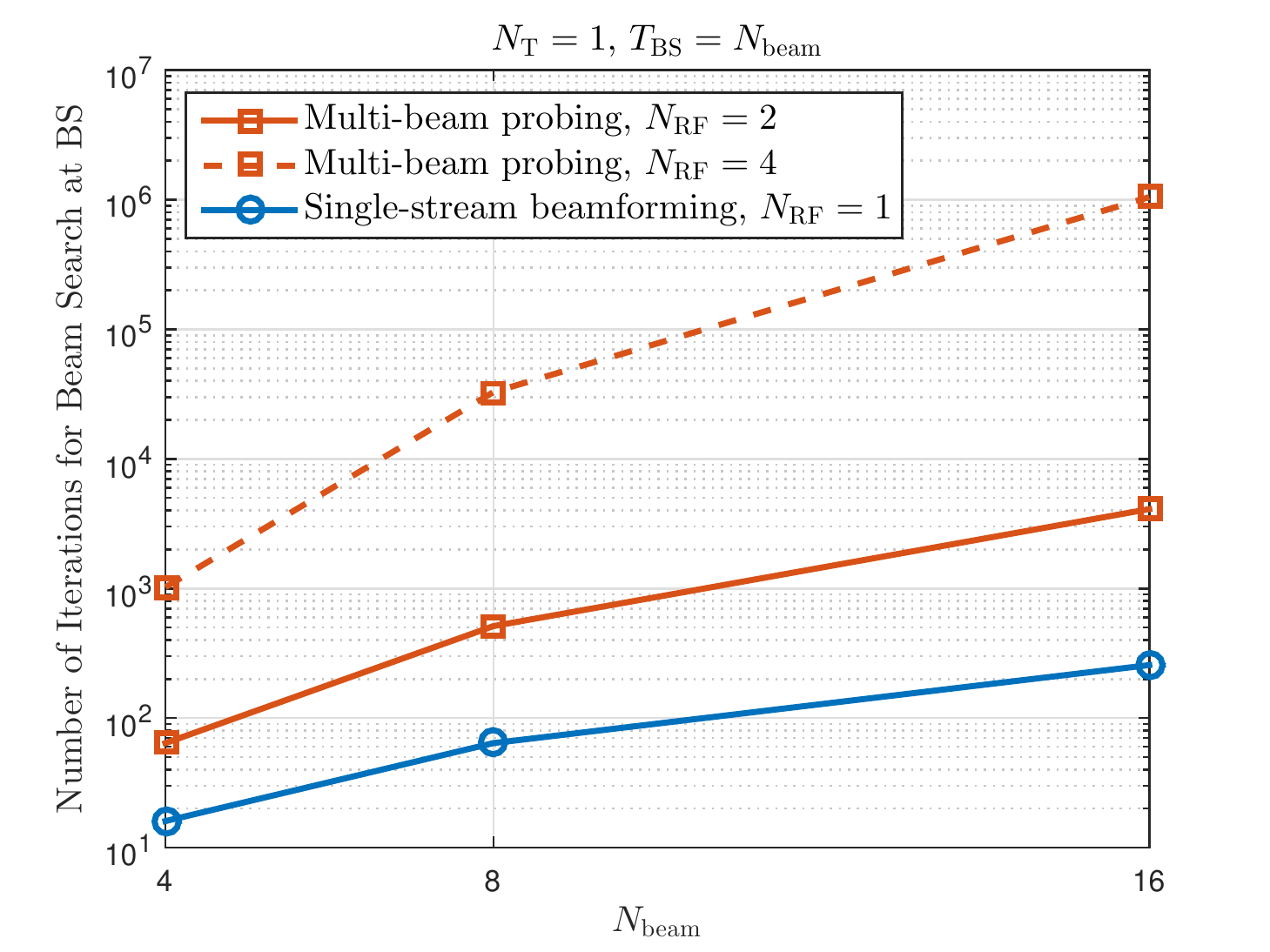}
\label{fig:subfigurecomplexa}}
%\hspace{-3.5mm}
\subfigure[]{%
\includegraphics[width=2.6in]{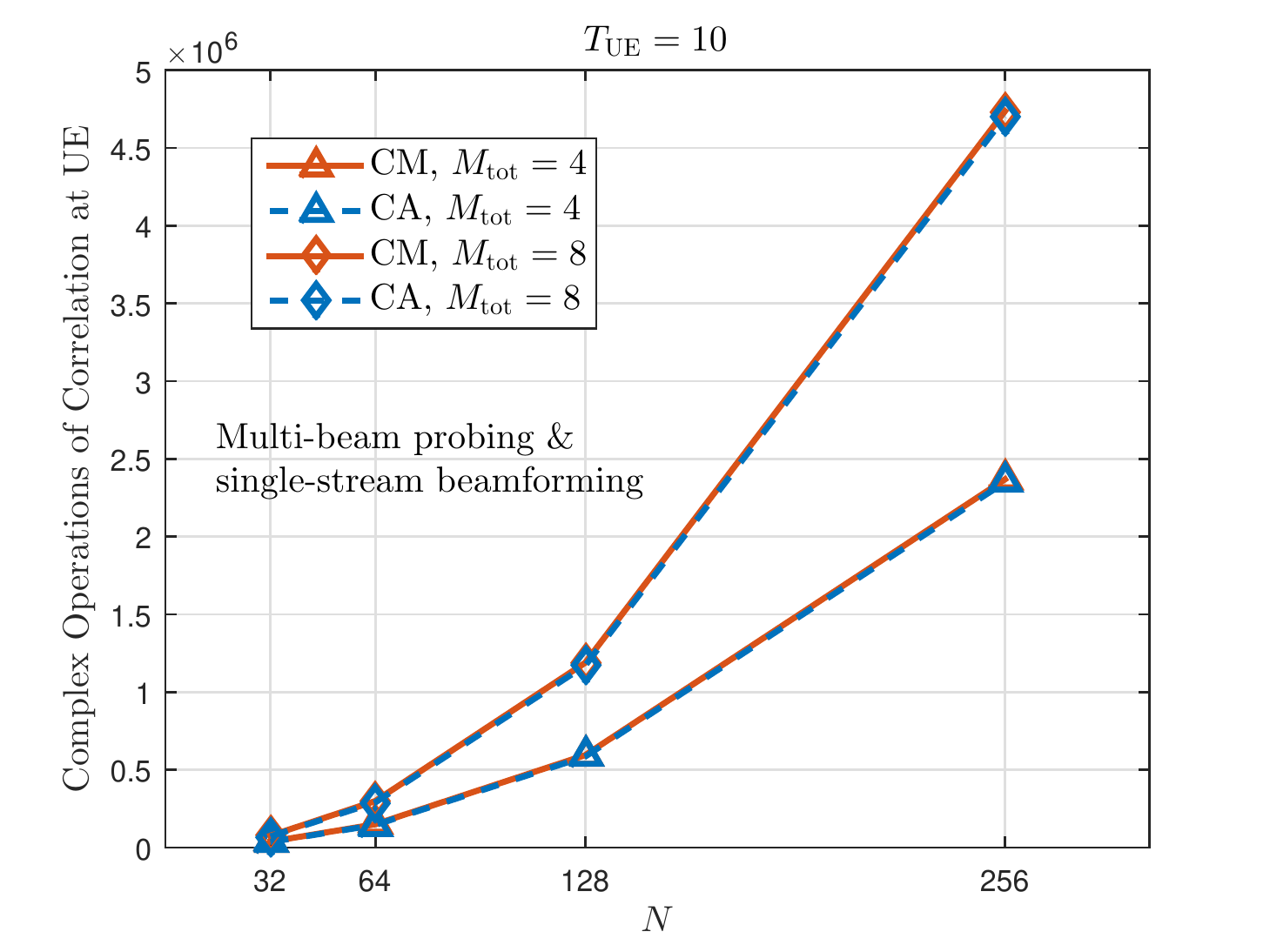}
\label{fig:subfigurecomplexb}}
\caption{(a) Beam search complexity at the BS in terms of the number of iterations over a given beam codebook (with $N_{\mathrm{beam}}$ beam codewords) across all $N_{\mathrm{RF}}$ transmit RF chains. (b) The number of complex multiplication (CM) and complex addition (CA) operations required at the UE to perform the correlation. A total of $T_{\mathrm{UE}}=10$ symbols are assumed by the UE with various numbers of receive antennas and occupied subcarriers.}
\label{fig:figurecomplexab}
\end{center}
\end{figure}

\begin{figure}
\begin{center}
\includegraphics[width=2.6in]{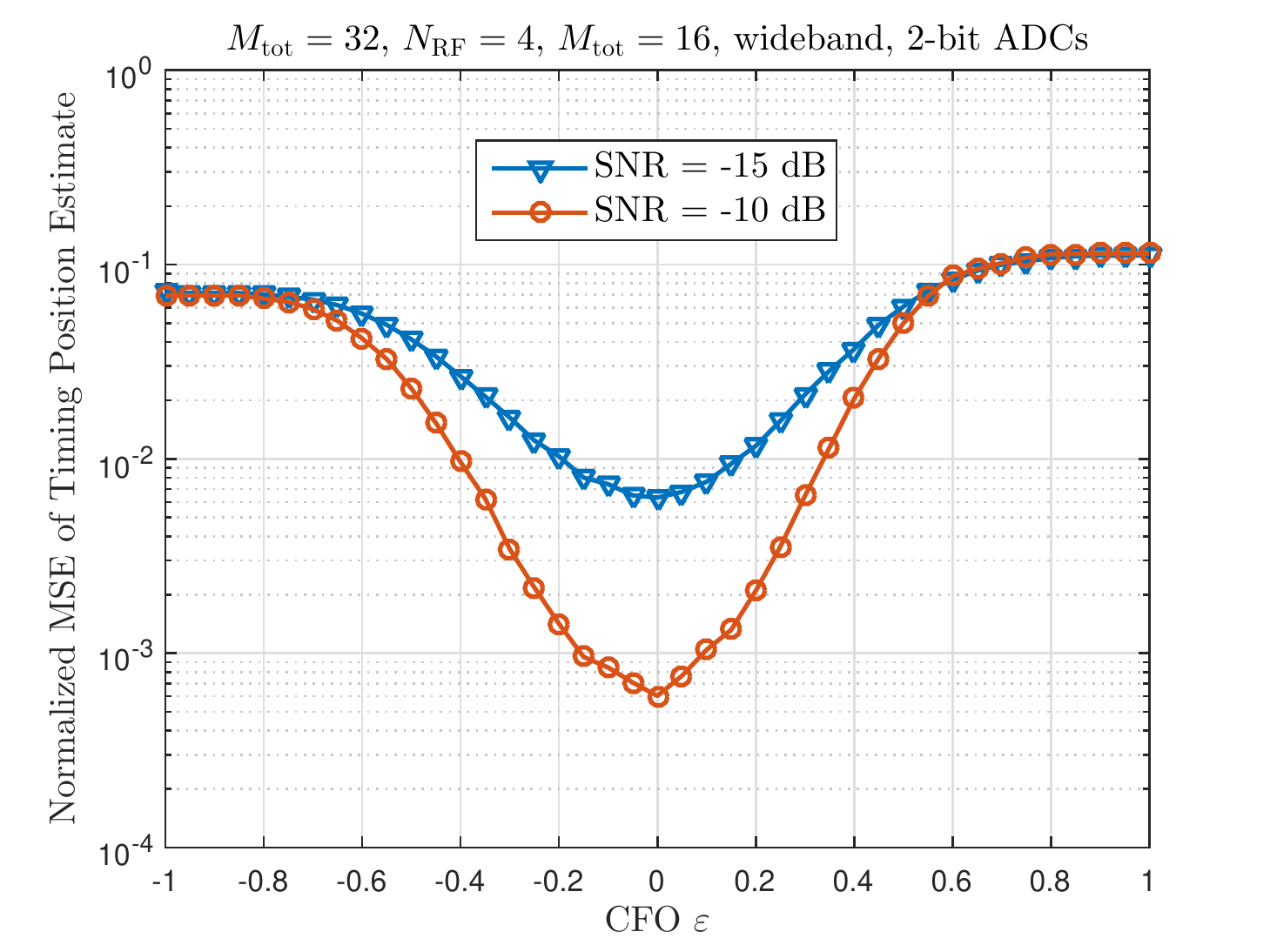}
\caption{Normalized MSE of timing position estimate versus carrier frequency offset. Multi-path frequency-selective channels are employed assuming a single UE equipped with a $2$-bit ADCs receiver; $N_{\mathrm{tot}}=32$ with $N_{\mathrm{RF}}=4$.}
\label{fig:figureCFO}
\end{center}
\end{figure}

\begin{figure}
\begin{center}
\subfigure[]{%
\includegraphics[width=2.6in]{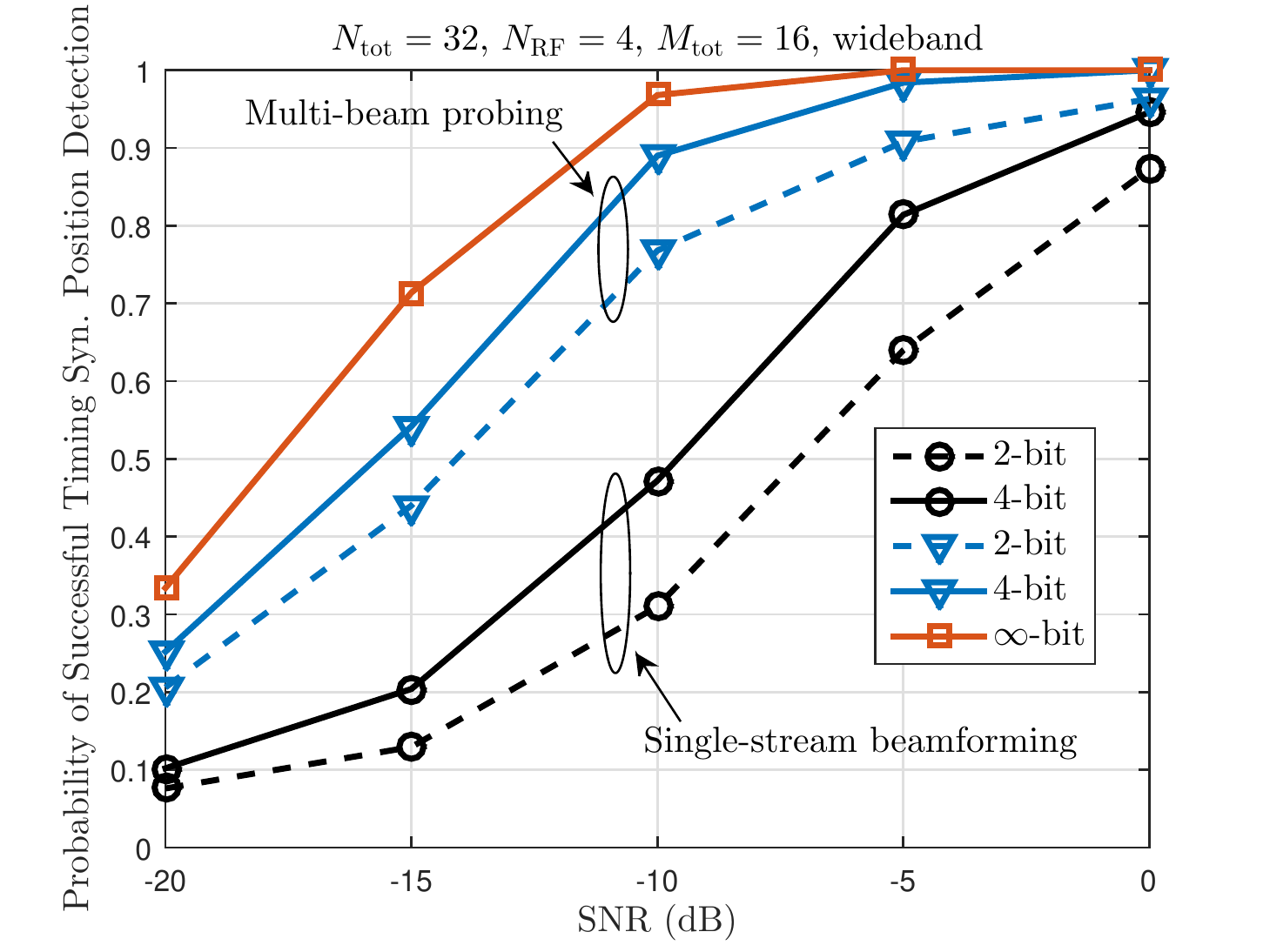}
\label{fig:figure10}}
%\hspace{-3.5mm}
\subfigure[]{%
\includegraphics[width=2.6in]{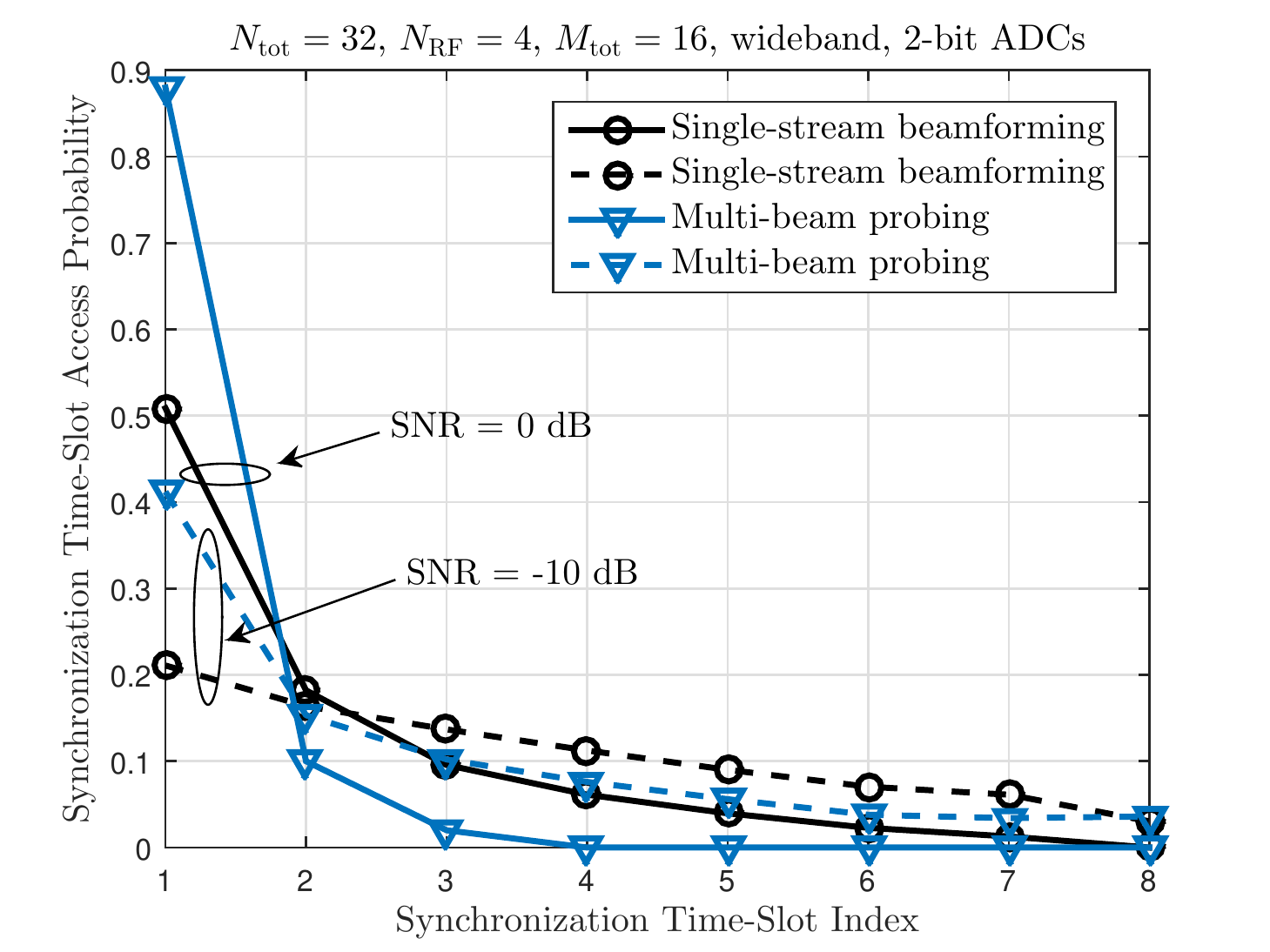}
\label{fig:figure11}}
\caption{(a) Probability of successful timing position detection versus SNR. A multi-cell scenario is assumed with a total of $7$ hexagonal cells. A number of $10$ UEs are randomly distributed within each cell sector with $2$ or $4$-bit ADCs. (b) Synchronization time-slot access probability is evaluated assuming a multi-cell scenario. A total of $7$ hexagonal cells are deployed. A number of $10$ UEs are randomly distributed within each cell sector with $2$-bit ADCs. The performance is evaluated under both $0$ dB and $-10$ dB SNRs.}
\label{fig:figureSpeciald}
\end{center}
\end{figure}

In this section, we evaluate the proposed multi-beam directional frame timing synchronization design for mmWave systems operating with low-resolution ADCs. The BS and UE employ a UPA and a ULA with inter-element spacing of $\lambda/2$ between the antenna elements. The BS covers three sectors, and each sector covers $120^{\circ}$ angular range $[-60^{\circ},60^{\circ}]$ around azimuth boresight ($0^{\circ}$) and $90^{\circ}$ angular range $[-45^{\circ},45^{\circ}]$ around elevation boresight ($0^{\circ}$). The UE monitors the entire $180^{\circ}$ angular region $[-90^{\circ},90^{\circ}]$ around boresight ($0^{\circ}$). We assume a $125$ MHz RF bandwidth with $N=512$ subcarriers. The corresponding CP length is $D=64$. We set the subcarrier spacing and symbol duration as $270$ KHz and $3.7$ $\mu s$ following the numerology provided in \cite{jerrypi}. The synchronization sequence occupies the central $63$ subcarriers, and we set the DC-carrier as zero. We set the total number of receive antennas $M_{\mathrm{tot}}$ as $16$ throughout the simulation section unless otherwise specified. In Figs.~\ref{fig:figure8ab}, \ref{fig:subfigure9a} and \ref{fig:figureCFO}, we assume a single UE, while we deploy multiple UEs in a single cell in Figs.~\ref{fig:subfigure9b} and \ref{fig:figure10}. In Fig.~\ref{fig:figure11}, we examine the proposed method in a multi-cell scenario. We evaluate the implementation complexity in Fig.~\ref{fig:figurecomplexab} and the impact of the CFO in Fig.~\ref{fig:figureCFO}. For the multi-user setup, we let $1/\lambda'_{\mathrm{max}}$ be $-20$ dB. Further, we assume that the ADCs equipped at all UEs have identical quantization resolution ($2$ or $4$ bits), which is used to determine $\xi_{\mathrm{max}}$.
%according to \cite[Table~I]{chestjm}.

In Fig.~\ref{fig:subfigure8a}, we plot the cumulative density functions (CDFs) of the received synchronization SQNRs at zero-lag correlation for $N_{\mathrm{tot}}=32$ with $N_{\mathrm{RF}}=4$. We assume single-path frequency-flat channels. We set the transmit SNR as $0$ dB, which is calculated before the transmit beamforming and receive processing. We also examine the single-stream beamforming based directional frame timing synchronization method described in Section II with infinite-resolution ($\infty$) and low-resolution ($2$ and $4$ bits) ADCs for comparison. We construct the employed synchronization signals according to (\ref{zcfdd}) with root index $34$. As can be seen from Fig.~\ref{fig:subfigure8a}, a significant performance gap can be observed between low-resolution quantization and infinite-resolution quantization for the existing strategy, though with increase in the quantization resolution, this performance difference decreases. By using the proposed multi-beam probing based design approach, the received synchronization SQNR performance is improved by several orders of decibels in contrast to the single-stream beamforming algorithm. In Fig.~\ref{fig:subfigure8b}, we provide the average received synchronization SQNRs at zero-lag correlation versus the quantization resolution of ADCs. In this example, we implement the statistical mmWave wideband channel model developed in \cite{kssr} using the NYUSIM open source platform. The urban micro-cellular (UMi) scenario is assumed with non-line-of-sight (NLOS) components for $28$ GHz carrier frequency. We evaluate both $N_{\mathrm{tot}}=32$ and $N_{\mathrm{tot}}=64$ for the proposed multi-beam probing and conventional single-stream beamforming based designs. Similar to the narrowband results shown in Fig.~\ref{fig:subfigure8a}, the proposed algorithm exhibits superior synchronization SQNR performance over the single-stream beamforming based design in wideband channels.
%The channel is modeled as a clustered channel, where each cluster comprises several subpaths. Detailed channel modeling parameters including the distributions of clusters, subpaths in each cluster, and etc. are given in \cite[TABLE~III]{kssr}.

In Fig.~\ref{fig:subfigure9a}, we evaluate the normalized MSE (NMSE) performance of the timing position estimate assuming a single UE. Denoting the NMSE of the timing position estimate by $\omega_{\mathrm{est}}$, we have
\begin{equation}
\omega_{\mathrm{est}} = \mathbb{E}\left[\left|\frac{\kappa_{\mathrm{true}}-\kappa_{\mathrm{est}}}{\kappa_{\mathrm{true}}}\right|^{2}\right],
\end{equation}
where $\kappa_{\mathrm{true}}$ denotes the first sample index of the probed synchronization signal, and $\kappa_{\mathrm{est}}$ corresponds to its estimated counterpart using various synchronization methods. We employ the wideband channels with $2$-bit ADCs. It can be observed from Fig.~\ref{fig:subfigure9a} that the performance gap between low-resolution quantization and infinite-resolution quantization is significant for the existing method especially at relatively high SNR. The proposed multi-beam directional synchronization method with common synchronization signal design approaches the infinite-resolution case for various SNR values. In Fig.~\ref{fig:subfigure9b}, we examine the NMSE performance of the timing position estimate in a single-cell multi-user scenario. In this example, we randomly drop a total of $10$ UEs within a circular cell having a $150$ $m$ radius. We employ the same path loss model as in \cite{kssr}. In this example, we obtain the NMSE performance of the timing position estimate over all UEs. Similar observations as in Fig.~\ref{fig:subfigure9a} can be obtained, i.e., the proposed method exhibits better performance than the existing directional timing synchronization design.

In Fig.~\ref{fig:subfigurecomplexa}, we plot the number of iterations required at the BS for the proposed multi-beam probing and conventional single-stream beamforming based synchronization designs. As can be seen from Fig.~\ref{fig:subfigurecomplexa}, the beam search complexity of our proposed approach is significantly larger than that of the conventional strategy especially for relatively large $N_{\mathrm{RF}}$ ($4$ in this example). In practice, however, the BS would only execute the beam search/optimization in a semi-static manner. In Fig.~\ref{fig:subfigurecomplexb}, we calculate the number of complex multiplication (CM) and complex addition (CA) operations required at the UE when conducting the correlation based timing synchronization. From the plot, it can be observed that the proposed multi-beam probing based strategy does not incorporate additional implementation complexity at the UE.

In Fig.~\ref{fig:figureCFO}, we examine the CFO impact on the frame timing synchronization performance of the proposed multi-beam probing strategy. The simulation setup in this example is identical to that in Fig.~\ref{fig:subfigure9a}. As can be seen from Fig.~\ref{fig:figureCFO}, as the CFO $\varepsilon$ increases, the NMSE performance of the timing position estimate degrades for both $-10$ dB and $-15$ dB SNRs. However, even under relatively large CFOs, e.g., $\varepsilon=+1$ and $-1$, the proposed approach still exhibits similar timing synchronization performance relative to the conventional method without CFO.

For the last two examples, we numerically study the proposed method in a multi-cell scenario. Specifically, we assume $7$ hexagonal cells and set the inter-site distance as $500$ $m$. The root indices of the employed ZC sequences are $25$, $29$ and $34$. The central cell corresponds to root index $25$, and root indices $29$ and $34$ are reused among the other $6$ surrounding cells such that the neighboring two cells are assigned two distinct root indices. Similar to the example provided in Fig.~\ref{fig:subfigure9b}, we randomly drop a total of $10$ UEs within each cell sector, and we set the minimum distance between the BS and UE as $20$ $m$. Performance statistics are obtained from the central cell of interest only. In Fig.~\ref{fig:figure10}, we evaluate the probability of successful timing position detection. Different from the NMSE of the timing position estimate, we calculate the probability of successful detection as $\mathrm{Pr}\left(\kappa_{\mathrm{est}}=\kappa_{\mathrm{true}}\right)$. It is observed that even at relatively low SNR, the proposed synchronization method with low-resolution ADCs can still achieve promising detection performance. For instance, at $-10$ dB SNR, the probability of successful detection is more than $0.8$ for the proposed multi-beam probing with common synchronization signal design.

In Fig.~\ref{fig:figure11}, we evaluate the synchronization time-slot access probability under low-resolution quantization. We calculate the synchronization time-slot access probability based on the following procedure. If a given UE successfully detects the frame timing position during synchronization time-slot $\tau_{\mathrm{t}}$, the algorithm stops for the UE of interest. Otherwise, the algorithm continues in synchronization time-slot $\tau_{\mathrm{t}}+1$ and repeats the previous procedure. The synchronization time-slot access probability evaluates how fast a specific UE is capable of correctly detecting the frame timing position, which in turn implies the access delay for the given UE. Fig.~\ref{fig:figure11} shows that even with $2$-bit ADCs, the UEs can correctly detect the frame timing position during the first several synchronization time-slots by using the proposed method.
%The results shown in Figs.~\ref{fig:figure9ab}$\sim$\ref{fig:figureSpeciald} reveal that optimizing the received synchronization SQNR at zero-lag correlation would improve the low-resolution frame timing synchronization performances. This observation is consistent with our analysis in Sections III and IV.
\section{Conclusions}
In this paper, we developed and evaluated a multi-beam probing-assisted directional frame timing synchronization method for mmWave systems with low-resolution ADCs. We first formulated the optimization problem as maximizing the minimum received synchronization SQNR at zero-lag correlation among all UEs. We solved this problem by transforming the complex max-min multicast beamforming problem into a maximization problem. We then proposed a multi-beam probing strategy to tackle the maximization problem by optimizing the effective composite beam pattern. We showed via numerical examples that the effective composite beam can provide a good tradeoff between the beamforming gain and the quantization distortion and characterize the worst-case scenario of the network.
%The proposed method achieves promising frame timing synchronization performance and reduces the access delay even under $2$-bit quantization resolution.
\appendix[Proof of Lemma 1]
We first compute the non-zero-lag (lag $\upsilon$) frequency-domain correlation between the received signal samples and the known unquantized reference synchronization sequence $\textbf{\textsf{d}}$ as
\begin{eqnarray}\label{nonzerolagfreq}
\Lambda^{(0)}_{u,\hat{b}}[\upsilon] &=& \sum_{k=0}^{N-1}\textsf{q}^{(0)}_{u,\hat{b}}[k+\upsilon]\textsf{d}^{*}[k]\\
&=&\sum_{k=0}^{N-1}\eta^{(0)}_{u,\hat{b}}[k+\upsilon]\left[\bm{A}^{\mathrm{RX}}_{u}\textbf{\textsf{G}}_u[k+\upsilon]\left(\bm{A}^{\mathrm{TX}}_{u}\right)^{*}\right]_{\hat{b},:}\bm{f}_{0}\textsf{d}[k+\upsilon]\textsf{d}^{*}[k]\nonumber\\ \label{llllrr22}
&+&\sum_{k=0}^{N-1}\eta^{(0)}_{u,\hat{b}}[k+\upsilon]\textsf{w}_u[k+\upsilon]\textsf{d}^{*}[k]+\sum_{k=0}^{N-1}\check{\textsf{w}}^{(0)}_{u,\hat{b}}[k+\upsilon]\textsf{d}^{*}[k].
\end{eqnarray}
Assuming flat synchronization channels, we can rewrite (\ref{llllrr22}) as
\begin{eqnarray}\label{nonzerolagfreqsingle}
\Lambda^{(0)}_{u,\hat{b}}[\upsilon] &=&\underline{\eta}^{(0)}_{u,\hat{b}}\textsf{g}_{u}\left[\bm{a}_{\mathrm{rx}}(\psi_u)\bm{a}^{*}_{\mathrm{tx}}(\theta_u,\phi_u)\right]_{\hat{b},:}\bm{f}_{0}\sum_{k=0}^{N-1}\textsf{d}[k+\upsilon]\textsf{d}^{*}[k]\nonumber\\
&+&\underline{\eta}^{(0)}_{u,\hat{b}}\sum_{k=0}^{N-1}\textsf{w}_u[k+\upsilon]\textsf{d}^{*}[k]+\sum_{k=0}^{N-1}\check{\textsf{w}}^{(0)}_{u,\hat{b}}[k+\upsilon]\textsf{d}^{*}[k].
\end{eqnarray}
Based upon (\ref{fcorrex}) and (\ref{dzzzzzz}), we can obtain
\begin{eqnarray}\label{nzlnzl}
\varsigma^{\mathrm{nzl},(0)}_{u,\hat{b}}&=&\mathbb{E}\left[|\Lambda^{(0)}_{u,\hat{b}}[\upsilon]|^{2}\right]\\
&=&\left(\underline{\eta}^{(0)}_{u,\hat{b}}\right)^2\sigma^{2}+\underline{\eta}^{(0)}_{u,\hat{b}}\left(1-\underline{\eta}^{(0)}_{u,\hat{b}}\right)\left(\textsf{g}_{u}^{2}\left|\left[\bm{a}_{\mathrm{rx}}(\psi_u)\bm{a}^{*}_{\mathrm{tx}}(\theta_u,\phi_u)\right]_{\hat{b},:}\bm{f}_{0}\right|^2+\sigma^2\right).
\end{eqnarray}
Similar to (\ref{nzlnzl}), we can compute $\varsigma^{\mathrm{zl},(0)}_{u,\hat{b}}$ for the zero-lag correlation as
\begin{eqnarray}\label{zlzlzl}
\varsigma^{\mathrm{zl},(0)}_{u,\hat{b}}&=&\mathbb{E}\left[|\Lambda^{(0)}_{u,\hat{b}}[0]|^{2}\right]\\
&=&\left(\underline{\eta}^{(0)}_{u,\hat{b}}\right)^2\textsf{g}_{u}^{2}\left|\left[\bm{a}_{\mathrm{rx}}(\psi_u)\bm{a}^{*}_{\mathrm{tx}}(\theta_u,\phi_u)\right]_{\hat{b},:}\bm{f}_{0}\right|^2\nonumber\\
&+&\left(\underline{\eta}^{(0)}_{u,\hat{b}}\right)^2\sigma^{2}+\underline{\eta}^{(0)}_{u,\hat{b}}\left(1-\underline{\eta}^{(0)}_{u,\hat{b}}\right)\left(\textsf{g}_{u}^{2}\left|\left[\bm{a}_{\mathrm{rx}}(\psi_u)\bm{a}^{*}_{\mathrm{tx}}(\theta_u,\phi_u)\right]_{\hat{b},:}\bm{f}_{0}\right|^2+\sigma^2\right).
\end{eqnarray}
We can therefore calculate the power ratio $\varsigma^{(0)}_{u,\hat{b}}$ as
\begin{eqnarray}\label{zlnzlratio}
\varsigma^{(0)}_{u,\hat{b}} = \frac{\varsigma^{\mathrm{zl},(0)}_{u,\hat{b}}}{\varsigma^{\mathrm{nzl},(0)}_{u,\hat{b}}}=1+\gamma^{(0)}_{u,\hat{b}},
\end{eqnarray}
which completes the proof.
%\appendix[Common definitions]
%\begin{table}

%\end{table}
\bibliographystyle{IEEEbib}
\bibliography{main_bib_TWC}
\end{document}